
\input harvmac
\input amssym



\newfam\frakfam
\font\teneufm=eufm10
\font\seveneufm=eufm7
\font\fiveeufm=eufm5
\textfont\frakfam=\teneufm
\scriptfont\frakfam=\seveneufm
\scriptscriptfont\frakfam=\fiveeufm


\def\bb{
\font\tenmsb=msbm10
\font\sevenmsb=msbm7
\font\fivemsb=msbm5
\textfont1=\tenmsb
\scriptfont1=\sevenmsb
\scriptscriptfont1=\fivemsb
}



\newfam\dsromfam
\font\tendsrom=dsrom10
\textfont\dsromfam=\tendsrom
\def\ds{\fam\dsromfam \tendsrom}


\newfam\mbffam
\font\tenmbf=cmmib10
\font\sevenmbf=cmmib7
\font\fivembf=cmmib5
\textfont\mbffam=\tenmbf
\scriptfont\mbffam=\sevenmbf
\scriptscriptfont\mbffam=\fivembf


\newfam\mbfcalfam
\font\tenmbfcal=cmbsy10
\font\sevenmbfcal=cmbsy7
\font\fivembfcal=cmbsy5
\textfont\mbfcalfam=\tenmbfcal
\scriptfont\mbfcalfam=\sevenmbfcal
\scriptscriptfont\mbfcalfam=\fivembfcal


\newfam\mscrfam
\font\tenmscr=rsfs10
\font\sevenmscr=rsfs7
\font\fivemscr=rsfs5
\textfont\mscrfam=\tenmscr
\scriptfont\mscrfam=\sevenmscr
\scriptscriptfont\mscrfam=\fivemscr
\def\scr{\fam\mscrfam \tenmscr}




\def\tilde{\widetilde}
\def\t{\tilde}
\def\hat{\widehat}

\def\bar{\overline}
\def\b{\bar}
\def\bsq#1{{{\b{#1}}^{\lower 2.5pt\hbox{$\scriptstyle 2$}}}}
\def\bexp#1#2{{{\b{#1}}^{\lower 2.5pt\hbox{$\scriptstyle #2$}}}}
\def\dotexp#1#2{{{#1}^{\lower 2.5pt\hbox{$\scriptstyle #2$}}}}


\def\rt2{\sqrt{2}}
\def\half {{1 \over 2}}

\def\d{\partial}

\def\det{\mathop{\rm det}}

\def\Tr{\mathop{\rm Tr}}


\font\tenbifull=cmmib10
\font\tenbimed=cmmib7
\font\tenbismall=cmmib5
\textfont9=\tenbifull \scriptfont9=\tenbimed
\scriptscriptfont9=\tenbismall

\mathchardef\bbGamma="7000
\mathchardef\bbDelta="7001
\mathchardef\bbPhi="7002
\mathchardef\bbAlpha="7003
\mathchardef\bbXi="7004
\mathchardef\bbPi="7005
\mathchardef\bbSigma="7006
\mathchardef\bbUpsilon="7007
\mathchardef\bbTheta="7008
\mathchardef\bbPsi="7009
\mathchardef\bbOmega="700A
\mathchardef\bbalpha="710B
\mathchardef\bbbeta="710C
\mathchardef\bbgamma="710D
\mathchardef\bbdelta="710E
\mathchardef\bbepsilon="710F
\mathchardef\bbzeta="7110
\mathchardef\bbeta="7111
\mathchardef\bbtheta="7112
\mathchardef\bbiota="7113
\mathchardef\bbkappa="7114
\mathchardef\bblambda="7115
\mathchardef\bbmu="7116
\mathchardef\bbnu="7117
\mathchardef\bbxi="7118
\mathchardef\bbpi="7119
\mathchardef\bbrho="711A
\mathchardef\bbsigma="711B
\mathchardef\bbtau="711C
\mathchardef\bbupsilon="711D
\mathchardef\bbphi="711E
\mathchardef\bbchi="711F
\mathchardef\bbpsi="7120
\mathchardef\bbomega="7121
\mathchardef\bbvarepsilon="7122
\mathchardef\bbvartheta="7123
\mathchardef\bbvarpi="7124
\mathchardef\bbvarrho="7125
\mathchardef\bbvarsigma="7126
\mathchardef\bbvarphi="7127


\def\alphadot{{\dot\alpha}}
\def\betadot{{\dot\beta}}


\def\ibar{\b{i}}
\def\jbar{\b{j}}


\def\thetabar{\b{\theta}}
\def\thetasq{\theta^2}
\def\thetabarsq{{\thetabar^2}}

\def\jbar{\b{j}}


\def\CA{{\cal A}}

\def\CG{{\cal G}}

\def\CJ{{\cal J}}
\def\CK{{\cal K}}

\def\CN{{\cal N}}
\def\CO{{\cal O}}

\def\CQ{{\cal Q}}
\def\CR{{\cal R}}
\def\CS{{\cal S}}
\def\CT{{\cal T}}

\def\CW{{\cal W}}

\def\CY{{\cal Y}}
\def\CZ{{\cal Z}}


\def\1{{\ds 1}}
\def\R{\hbox{$\bb R$}}
\def\C{\hbox{$\bb C$}}

\def\Z{\hbox{$\bb Z$}}

\def\P{\hbox{$\bb P$}}


\def\ep{\varepsilon}

\noblackbox




\lref\WessCP{
  J.~Wess, J.~Bagger,
  ``Supersymmetry and Supergravity,''
Princeton, Univ. Pr. (1992).
}

\lref\deAzcarragaGM{
  J.~A.~de Azcarraga, J.~P.~Gauntlett, J.~M.~Izquierdo, P.~K.~Townsend,
  ``Topological Extensions of the Supersymmetry Algebra for Extended Objects,''
Phys.\ Rev.\ Lett.\  {\bf 63}, 2443 (1989).
}

\lref\FerraraTX{
  S.~Ferrara, M.~Porrati,
  ``Central extensions of supersymmetry in four-dimensions and three-dimensions,''
Phys.\ Lett.\  {\bf B423}, 255-260 (1998).
[hep-th/9711116].
}

\lref\GorskyHK{
  A.~Gorsky, M.~A.~Shifman,
  ``More on the tensorial central charges in N=1 supersymmetric gauge theories (BPS wall junctions and strings),''
Phys.\ Rev.\  {\bf D61}, 085001 (2000).
[hep-th/9909015].
}

\lref\FerraraPZ{
  S.~Ferrara, B.~Zumino,
  ``Transformation Properties of the Supercurrent,''
Nucl.\ Phys.\  {\bf B87}, 207 (1975).
}

\lref\ClarkJX{
  T.~E.~Clark, O.~Piguet, K.~Sibold,
  ``Supercurrents, Renormalization and Anomalies,''
Nucl.\ Phys.\  {\bf B143}, 445 (1978).
}

\lref\GatesYC{
  S.~J.~Gates, Jr., M.~T.~Grisaru, W.~Siegel,
  ``Auxiliary Field Anomalies,''
Nucl.\ Phys.\  {\bf B203}, 189 (1982).
}

\lref\ShifmanZI{
  M.~A.~Shifman, A.~I.~Vainshtein,
  ``Solution of the Anomaly Puzzle in SUSY Gauge Theories and the Wilson Operator Expansion,''
Nucl.\ Phys.\  {\bf B277}, 456 (1986).
}

\lref\MagroAJ{
  M.~Magro, I.~Sachs, S.~Wolf,
  ``Superfield Noether procedure,''
Annals Phys.\  {\bf 298}, 123-166 (2002).
[hep-th/0110131].
}

\lref\KomargodskiPC{
  Z.~Komargodski, N.~Seiberg,
  ``Comments on the Fayet-Iliopoulos Term in Field Theory and Supergravity,''
JHEP {\bf 0906}, 007 (2009).
[arXiv:0904.1159 [hep-th]].
}

\lref\DienesTD{
  K.~R.~Dienes, B.~Thomas,
  ``On the Inconsistency of Fayet-Iliopoulos Terms in Supergravity Theories,''
Phys.\ Rev.\  {\bf D81}, 065023 (2010).
[arXiv:0911.0677 [hep-th]].
}

\lref\KuzenkoYM{
  S.~M.~Kuzenko,
  ``The Fayet-Iliopoulos term and nonlinear self-duality,''
Phys.\ Rev.\  {\bf D81}, 085036 (2010).
[arXiv:0911.5190 [hep-th]].
}

\lref\KomargodskiRB{
  Z.~Komargodski, N.~Seiberg,
  ``Comments on Supercurrent Multiplets, Supersymmetric Field Theories and Supergravity,''
JHEP {\bf 1007}, 017 (2010).
[arXiv:1002.2228 [hep-th]].
}

\lref\KuzenkoAM{
  S.~M.~Kuzenko,
  ``Variant supercurrent multiplets,''
JHEP {\bf 1004}, 022 (2010).
[arXiv:1002.4932 [hep-th]].
}

\lref\ZhengXX{
  S.~Zheng, J.~-h.~Huang,
  ``Variant Supercurrents and Linearized Supergravity,''
Class.\ Quant.\ Grav.\  {\bf 28}, 075012 (2011).
[arXiv:1007.3092 [hep-th]].
}

\lref\KuzenkoNI{
  S.~M.~Kuzenko,
  ``Variant supercurrents and Noether procedure,''
[arXiv:1008.1877 [hep-th]].
}

\lref\GatesNR{
  S.~J.~Gates, M.~T.~Grisaru, M.~Rocek, W.~Siegel,
  ``Superspace Or One Thousand and One Lessons in Supersymmetry,''
Front.\ Phys.\  {\bf 58}, 1-548 (1983).
[hep-th/0108200].
}

\lref\BanksZN{
  T.~Banks, N.~Seiberg,
  ``Symmetries and Strings in Field Theory and Gravity,''
[arXiv:1011.5120 [hep-th]].
}

\lref\HughesDN{
  J.~Hughes, J.~Polchinski,
  ``Partially Broken Global Supersymmetry and the Superstring,''
Nucl.\ Phys.\  {\bf B278}, 147 (1986).
}

\lref\DvaliXE{
  G.~R.~Dvali, M.~A.~Shifman,
  ``Domain walls in strongly coupled theories,''
Phys.\ Lett.\  {\bf B396}, 64-69 (1997).
[hep-th/9612128].
}

\lref\GauntlettCH{
  J.~P.~Gauntlett, G.~W.~Gibbons, C.~M.~Hull, P.~K.~Townsend,
  ``BPS states of D = 4 N=1 supersymmetry,''
Commun.\ Math.\ Phys.\  {\bf 216}, 431-459 (2001).
[hep-th/0001024].
}

\lref\ShifmanZZ{
  M.~Shifman, A.~Yung,
  ``Supersymmetric solitons,''
Cambridge, Univ. Pr. (2009).
}

\lref\RitzMP{
  A.~Ritz, M.~Shifman, A.~Vainshtein,
  ``Enhanced worldvolume supersymmetry and intersecting domain walls in N=1 SQCD,''
Phys.\ Rev.\  {\bf D70}, 095003 (2004).
[hep-th/0405175].
}

\lref\ChibisovRC{
  B.~Chibisov, M.~A.~Shifman,
  ``BPS saturated walls in supersymmetric theories,''
Phys.\ Rev.\  {\bf D56}, 7990-8013 (1997).
[hep-th/9706141].
}

\lref\VainshteinHU{
  A.~I.~Vainshtein, A.~Yung,
  ``Type I superconductivity upon monopole condensation in Seiberg-Witten theory,''
Nucl.\ Phys.\  {\bf B614}, 3-25 (2001).
[hep-th/0012250].
}

\lref\DavisBS{
  S.~C.~Davis, A.~-C.~Davis, M.~Trodden,
  ``N=1 supersymmetric cosmic strings,''
Phys.\ Lett.\  {\bf B405}, 257-264 (1997).
[hep-ph/9702360].
}

\lref\WittenPX{
  E.~Witten,
  ``Two-dimensional models with (0,2) supersymmetry: Perturbative aspects,''
[hep-th/0504078].
}

\lref\TanMI{
  M.~-C.~Tan, J.~Yagi,
  ``Chiral Algebras of (0,2) Sigma Models: Beyond Perturbation Theory,''
Lett.\ Math.\ Phys.\  {\bf 84}, 257-273 (2008).
[arXiv:0801.4782 [hep-th], arXiv:0805.1410 [hep-th]].
}

\lref\HouRV{
  X.~-r.~Hou, A.~Losev, M.~A.~Shifman,
  ``BPS saturated solitons in N=2 two-dimensional theories on R x S: domain walls in theories with compactified dimensions,''
Phys.\ Rev.\  {\bf D61}, 085005 (2000).
[hep-th/9910071].
}

\lref\BinosiWY{
  D.~Binosi, M.~A.~Shifman, T.~ter Veldhuis,
  ``Leaving the BPS bound: Tunneling of classically saturated solitons,''
Phys.\ Rev.\  {\bf D63}, 025006 (2001).
[hep-th/0006026].
}

\lref\WittenNF{
  E.~Witten,
  ``Dynamical Breaking of Supersymmetry,''
Nucl.\ Phys.\  {\bf B188}, 513 (1981).
}

\lref\BanksVH{
  T.~Banks, W.~Fischler, S.~H.~Shenker, L.~Susskind,
  ``M theory as a matrix model: A Conjecture,''
Phys.\ Rev.\  {\bf D55}, 5112-5128 (1997).
[hep-th/9610043].
}

\lref\BanksNN{
  T.~Banks, N.~Seiberg, S.~H.~Shenker,
  ``Branes from matrices,''
Nucl.\ Phys.\  {\bf B490}, 91-106 (1997).
[hep-th/9612157].
}

\lref\AchucarroQB{
  A.~Achucarro, J.~P.~Gauntlett, K.~Itoh, P.~K.~Townsend,
  ``World Volume Supersymmetry From Space-time Supersymmetry Of The Four-dimensional Supermembrane,''
Nucl.\ Phys.\  {\bf B314}, 129 (1989).
}

\lref\LosevGS{
  A.~Losev, M.~Shifman,
  ``N=2 sigma model with twisted mass and superpotential: Central charges and solitons,''
Phys.\ Rev.\  {\bf D68}, 045006 (2003).
[hep-th/0304003].
}

\lref\HughesFA{
  J.~Hughes, J.~Liu, J.~Polchinski,
  ``Supermembranes,''
Phys.\ Lett.\  {\bf B180}, 370 (1986).
}

\lref\AntoniadisVB{
  I.~Antoniadis, H.~Partouche, T.~R.~Taylor,
  ``Spontaneous breaking of N=2 global supersymmetry,''
Phys.\ Lett.\  {\bf B372}, 83-87 (1996).
[hep-th/9512006].
}

\lref\FerraraXI{
  S.~Ferrara, L.~Girardello, M.~Porrati,
  ``Spontaneous breaking of N=2 to N=1 in rigid and local supersymmetric theories,''
Phys.\ Lett.\  {\bf B376}, 275-281 (1996).
[hep-th/9512180].
}

\lref\BaggerWP{
  J.~Bagger, A.~Galperin,
  ``A New Goldstone multiplet for partially broken supersymmetry,''
Phys.\ Rev.\  {\bf D55}, 1091-1098 (1997).
[hep-th/9608177].
}

\lref\RocekHI{
  M.~Rocek and A.~A.~Tseytlin,
  ``Partial breaking of global D = 4 supersymmetry, constrained superfields,
  and three-brane actions,''
  Phys.\ Rev.\  D {\bf 59}, 106001 (1999)
  [arXiv:hep-th/9811232].
}

\lref\WittenUP{
E.~Witten, unpublished.
}

\lref\AffleckVC{
  I.~Affleck, M.~Dine, N.~Seiberg,
  ``Dynamical Supersymmetry Breaking In Chiral Theories,''
Phys.\ Lett.\  {\bf B137}, 187 (1984).
}

\lref\DineBK{
M.~Dine, ``Fields, Strings and Duality: TASI 96,'' eds. C. Efthimiou and B. Greene
(World Scientific, Singapore, 1997).}

\lref\WeinbergUV{
  S.~Weinberg,
  ``Nonrenormalization theorems in nonrenormalizable theories,''
Phys.\ Rev.\ Lett.\  {\bf 80}, 3702-3705 (1998).
[hep-th/9803099].
}

\lref\DumitrescuCA{
  T.~T.~Dumitrescu, Z.~Komargodski, M.~Sudano,
  ``Global Symmetries and D-Terms in Supersymmetric Field Theories,''
[arXiv:1007.5352 [hep-th]].
}

\lref\AbelWV{
  S.~Abel, M.~Buican, Z.~Komargodski,
  ``Mapping Anomalous Currents in Supersymmetric Dualities,''
[arXiv:1105.2885 [hep-th]].
}

\lref\AffleckMK{
  I.~Affleck, M.~Dine, N.~Seiberg,
  ``Dynamical Supersymmetry Breaking in Supersymmetric QCD,''
Nucl.\ Phys.\  {\bf B241}, 493-534 (1984).
}

\lref\SeibergBZ{
  N.~Seiberg,
  ``Exact results on the space of vacua of four-dimensional SUSY gauge theories,''
Phys.\ Rev.\  {\bf D49}, 6857-6863 (1994).
[hep-th/9402044].
}

\lref\SeibergPQ{
  N.~Seiberg,
  ``Electric - magnetic duality in supersymmetric nonAbelian gauge theories,''
Nucl.\ Phys.\  {\bf B435}, 129-146 (1995).
[hep-th/9411149].
}

\lref\IntriligatorAU{
  K.~A.~Intriligator, N.~Seiberg,
  ``Lectures on supersymmetric gauge theories and electric - magnetic duality,''
Nucl.\ Phys.\ Proc.\ Suppl.\  {\bf 45BC}, 1-28 (1996).
[hep-th/9509066].
}

\lref\SeibergVC{
  N.~Seiberg,
  ``Naturalness versus supersymmetric nonrenormalization theorems,''
Phys.\ Lett.\  {\bf B318}, 469-475 (1993).
[hep-ph/9309335].
}

\lref\CuiRZ{
  X.~Cui, M.~Shifman,
  ``N=(0,2) Supersymmetry and a Nonrenormalization Theorem,''
[arXiv:1105.5107 [hep-th]].
}

\lref\BelinfantePU{
 F.~J.~Belinfante,
 ``On the Current and the Density of the Electric Charge, the Energy, the Linear Momentum and the Angular Momentum of Arbitrary Fields,''
 Physica {\bf 7}, 449 (1940)
}

\lref\Brunnerqyf{
 I.~Brunner, J.~Schulz and A.~Tabler,
``Boundaries and supercurrent multiplets in 3D Landau-Ginzburg models,''
JHEP {\bf 06}, 046 (2019)
[arXiv:1904.07258 [hep-th]].
}


\rightline{PUPT-2372}
\Title{\vbox{\baselineskip2pt \hbox{}}} {\vbox{\centerline{Supercurrents and Brane Currents}
\vskip2pt
\centerline{in Diverse Dimensions}}}
\centerline{Thomas T. Dumitrescu$^1$ and Nathan Seiberg$^2$}
\bigskip
\centerline{$^1${\it Department of Physics, Princeton University, Princeton, NJ 08544, USA}}
\centerline{$^2${\it  School of Natural Sciences, Institute for Advanced Study, Princeton, NJ 08540, USA}}

\vskip35pt

\noindent
We systematically analyze all possible supersymmetry multiplets that include the supersymmetry current and the energy-momentum tensor in various dimensions, focusing on~$\CN=1$ in four dimensions.  The most general such multiplet is the~$\CS$-multiplet, which includes 16 bosonic and 16 fermionic operators.  In special situations it can be decomposed, leading to smaller multiplets with~$12+12$ or even~$8+8$ operators.  Physically, these multiplets give rise to different brane charges in the supersymmetry algebra.  The~$\CS$-multiplet is needed when the algebra contains both string and domain wall charges.  In lower dimensions (or in four-dimensional~$\CN=2$ theories) the algebra can include space-filling brane charges, which are associated with partial supersymmetry breaking.  This phenomenon is physically distinct from ordinary spontaneous supersymmetry breaking.  Our analysis leads to new results about the dynamics of supersymmetric field theories.  These include constraints on the existence of certain charged branes and the absence of magnetic charges in~$U(1)$ gauge theories with a Fayet-Iliopoulos term.

\vskip2.5cm

\Date{June 2011}


\newsec{Introduction}

The goal of this paper is to present a systematic analysis of~{\it supercurrents} -- supersymmetry (SUSY) multiplets that include the supersymmetry current and the energy-momentum tensor.  We find the most general consistent supercurrent and we show under what conditions it can be decomposed into smaller multiplets.  Furthermore, we give a physical interpretation of the various supercurrents.

For concreteness, we initially focus on~$\CN=1$ theories in four dimensions. Later we extend our discussion to~$\CN=2$ theories in three dimensions, as well as~$\CN=(0,2)$ and~$\CN=(2,2)$ theories in two dimensions.

In its simplest form, the~$\CN=1$ algebra in four dimensions is\foot{We follow the conventions of Wess and Bagger~\WessCP, except that our convention for switching between vectors and bi-spinors is
\eqn\bispve{\ell_{\alpha\alphadot} = -2 \sigma^\mu_{\alpha\alphadot} \ell_\mu~,\qquad \ell_\mu = {1 \over 4} \b \sigma_\mu^{\alphadot \alpha} \ell_{\alpha\alphadot}~.}}
\eqn\salg{\eqalign{& \{Q_\alpha, \b Q_\alphadot\} = 2 \sigma^\mu_{\alpha\alphadot} P_\mu~,\cr
& \{Q_\alpha, Q_\beta\} = 0~.}}
Following~\refs{\deAzcarragaGM\FerraraTX-\GorskyHK}, we can add additional charges~$Z_\mu$ and~$Z_{\mu\nu}$ to this algebra,
\eqn\extfdsalg{\eqalign{&\{Q_\alpha, \b Q_\alphadot\} = 2 \sigma^\mu_{\alpha\alphadot} \left( P_\mu + Z_\mu\right)~,\cr
& \{ Q_\alpha, Q_\beta\} = \sigma^{\mu\nu}_{\alpha\beta} Z_{\mu\nu}~.}}
The charges~$Z_\mu$ and~$Z_{\mu\nu}$ are {\it brane charges}.  They are nonzero for one-branes (strings) and two-branes (domain walls) respectively.  These brane charges commute with the supercharges, but they are not central charges of the super-Poincar\'e algebra, because they do not commute with the Lorentz generators.  Other known modifications of the supersymmetry algebra~\salg\ include terms that do not commute with the supercharges; we do not discuss them here.

The brane charges~$Z_\mu$ and~$Z_{\mu\nu}$ are generally infinite -- only the charge per unit volume is meaningful. This motivates us to replace the algebra~\extfdsalg\ by its local version,
\eqn\fdscurralgaa{\eqalign{&\{ \b Q_\alphadot, S_{\alpha \mu}\} = 2 \sigma^\nu_{\alpha\alphadot} \left(  T_{\nu\mu} +  C_{\nu\mu}\right) + \cdots~,\cr
& \{Q_\beta, S_{\alpha \mu}\} =   \sigma^{\nu\rho}_{\alpha\beta} C_{\nu\rho\mu} + \cdots~.}}
Here~$C_{\mu\nu}$ and~$C_{\mu\nu\rho}$ are {\it brane currents}.  They are the conserved currents corresponding to the brane charges~$Z_\mu$ and $Z_{\mu \nu}$. The SUSY current algebra~\fdscurralgaa\ implies that these brane currents are embedded in a supercurrent multiplet, along with the supersymmetry current~$S_{\alpha\mu}$ and the energy-momentum tensor~$T_{\mu\nu}$. The ellipses in~\fdscurralgaa\ represent Schwinger terms; they will be discussed below.

In general, both brane charges in~\extfdsalg\ are present and we must study the current algebra~\fdscurralgaa.  However, under certain conditions some of these charges are absent.  This means that the corresponding current is a total derivative.  In that case we should be able to set it to zero by an improvement transformation, which modifies the various operators in~\fdscurralgaa\ without affecting the associated charges. If this can be done, then the supercurrent multiplet contains fewer operators.

Supercurrent multiplets have been studied by many authors~\refs{\FerraraPZ \ClarkJX\GatesYC\ShifmanZI\MagroAJ\KomargodskiPC\DienesTD\KuzenkoYM\KomargodskiRB\KuzenkoAM\ZhengXX-\KuzenkoNI}; see also section~7.10 of~\GatesNR. Our discussion differs from earlier approaches in two crucial respects:
First, some authors view rigid supersymmetric field theory as a limit of a supergravity theory.  Supergravity has several known presentations, which differ in the choice of auxiliary as well as propagating fields.  These different supergravity theories are closely related to various supercurrents. We will pursue a complementary approach, focusing on the different supercurrent multiplets in the rigid theory. We then have the option of gauging these supercurrents to obtain a supergravity theory.  One advantage of this approach is that it can be used to derive constraints on consistent supergravity theories~\refs{\KomargodskiPC-\KuzenkoNI,\BanksZN}.

Second, we insist on discussing only well-defined operators.  These must be gauge invariant and globally well-defined, even when the target space of the theory has nontrivial topology.  It is sometimes useful to describe such well-defined operators in terms of other operators, which are not themselves well-defined.  A commonly known example arises in electrodynamics, where the field strength~$F_{\mu\nu}$ is gauge invariant and well-defined, but it is useful to express it in terms of the gauge non-invariant vector potential~$A_\mu$.  We will see that physically distinct supercurrent multiplets appear to be identical, if we are careless about allowing operators that are not well-defined.

Throughout our discussion of the various supercurrent multiplets, we impose the following basic requirements:
\medskip
\item{(a)} {\it The multiplet includes the energy-momentum tensor~$T_{\mu \nu}$.} Every local quantum field theory possesses a real, conserved, symmetric energy-momentum tensor (see appendix~A):
    \eqn\conseq{\d^\nu T_{\mu\nu} = 0~,\qquad  \qquad P_\mu = \int d^{D-1} x \,  {T_\mu}^0 ~.}
    The energy-momentum tensor is not unique. It can be modified by an improvement transformation
    \eqn\symimp{T_{\mu\nu} \rightarrow T_{\mu\nu} + \d_\mu U_\nu - \eta_{\mu\nu} \d^\rho U_\rho~, \qquad\qquad \d_{[\mu} U_{\nu]} = 0~.}
    More general improvement transformations include operators of higher spin; they will not be important for us.  The improvement term is automatically conserved, and it does not contribute to the total momentum $P_\mu$. The fact that~$U_\mu$ is closed ensures that~$T_{\mu\nu}$ remains symmetric. If there is a well-defined real scalar~$u$ such that~$U_\mu = \d_\mu u$, then the improvement~\symimp\ takes the more familiar form
    \eqn\symimpii{T_{\mu\nu} \rightarrow T_{\mu\nu} + \left(\d_\mu\d_\nu  - \eta_{\mu\nu} \d^2 \right)u~.}
\medskip
\item{(b)} {\it The multiplet includes the supersymmetry current~$S_{\alpha\mu}$.}  Every supersymmetric quantum field theory possesses a conserved supersymmetry current:
 \eqn\scurrcons{\d^\mu S_{\alpha \mu} = 0~, \qquad \qquad Q_\alpha = \int d^3x \,  {S_\alpha}^0~.}
    Like the energy-momentum tensor, the supersymmetry current is not unique. It can be modified by an improvement transformation
\eqn\scurrimp{S_{\alpha\mu} \rightarrow S_{\alpha\mu} + {\left(\sigma_{\mu\nu}\right)_\alpha}^\beta \d^\nu \omega_\beta~.}
    As before, more general improvements include operators of higher spin; we do not discuss them. The improvement term is automatically conserved and it does not affect the supercharges~$Q_\alpha$.
\medskip
\item{(c)} {\it The energy-momentum tensor and the supersymmetry current are the only operators with spin larger than one.}  This can be motivated by noting that when a rigid supersymmetric field theory is weakly coupled to supergravity, the supercurrent is the source of the metric superfield. Since the graviton and the gravitino are the only fields of spin larger than one in the supergravity multiplet, we demand that~$T_{\mu\nu}$ and~$S_{\alpha\mu}$ be the only operators of spin larger than one in the supercurrent.
\medskip
\item{(d)} {\it The multiplet is indecomposable.}  In other words, it cannot be separated into two decoupled supersymmetry multiplets.  This does not mean that the multiplet is irreducible.  As we will see below, most supercurrents are reducible -- they include a non-trivial sub-multiplet, which is closed under supersymmetry transformations.  However, if the complement of that sub-multiplet is not a separate supersymmetry multiplet, then the multiplet is indecomposable.

\bigskip

In section~2 we show that the most general supercurrent that satisfies the four basic requirements~(a)--(d) is a real superfield~$\CS_{\alpha\alphadot}$ obeying the constraints
\eqn\fdsmult{\eqalign{& \b D^\alphadot \CS_{\alpha\alphadot} = \chi_\alpha + \CY_\alpha~,\cr
& \b D_\alphadot \chi_\alpha = 0~,\qquad D^\alpha \chi_\alpha = \b D_\alphadot \b \chi^\alphadot~,\cr
& D_\alpha \CY_\beta + D_\beta \CY_\alpha = 0~,\qquad \b D^2 \CY_\alpha = 0~.}}
This multiplet must exist in every supersymmetric field theory. If there is a well-defined chiral superfield~$X$ such that~$\CY_\alpha = D_\alpha X$, the multiplet~\fdsmult\ reduces to the~$\CS$-multiplet of~\KomargodskiRB:
\eqn\smultks{\eqalign{& \b D^\alphadot \CS_{\alpha\alphadot} = \chi_\alpha + D_\alpha X~,\cr
& \b D_\alphadot \chi_\alpha = 0~,\qquad D^\alpha \chi_\alpha = \b D_\alphadot \b \chi^\alphadot~,\cr
& \b D_\alphadot X = 0~.}}
However, the superfield~$X$ does not always exist. Throughout this paper, we will refer to~\fdsmult\ as the~$\CS$-multiplet, and distinguish~\smultks\ as a special case.

The~$\CS$-multiplet is reducible (the superfields~$\chi_\alpha$ and~$\CY_\alpha$ are non-trivial sub-multiplets), but in general it is indecomposable. There are, however, special cases in which the~$\CS$-multiplet is decomposable, so that we can set either~$\chi_\alpha$, or~$\CY_\alpha$, or both to zero by an improvement transformation.  This gives rise to smaller supercurrent multiplets: the Ferrara-Zumino~(FZ) multiplet~\FerraraPZ\ with~$\chi_\alpha = 0$, the~$\CR$-multiplet~\refs{\GatesYC,\DienesTD\KuzenkoYM-\KomargodskiRB,\GatesNR} with~$\CY_\alpha = 0$, and the superconformal multiplet with~$\chi_\alpha = \CY_\alpha = 0$.

As an example, we discuss how the different supercurrent multiplets arise in general Wess-Zumino models.

In section~3 we analyze the supersymmetry current algebra \fdscurralgaa\ that follows from the different supercurrents discussed in section~2. We find that the difference between these multiplets is reflected in the brane currents they contain.

In section~4 we repeat the analysis of sections~2 and~3 for~$\CN = 2$ theories in three dimensions. We present the analogue of the~$\CS$-multiplet, and we explore the resulting current algebra to identify the different brane currents that can arise. As we will see, these theories admit space-filling brane currents, which are not present in four-dimensional theories with~$\CN=1$ supersymmetry.

In section~5 we discuss the~$\CS$-multiplet and the resulting current algebra in two-dimensional~$\CN=(0,2)$ theories.

In section~6 we present additional examples. In particular, we show that there are no magnetic charges in~$U(1)$ gauge theories with a Fayet-Iliopoulos (FI) term.

In section~7 we discuss partial supersymmetry breaking and its connection with space-filling brane currents. We show that these brane currents deform the supersymmetry current algebra by constants~\HughesDN. This highlights the fundamental qualitative difference between partial supersymmetry breaking and ordinary  spontaneous SUSY-breaking, where the current algebra is not modified.

In section~8 we consider the behavior of the supercurrent multiplet under renormalization group flow. This allows us to constrain the IR behavior of supersymmetric field theories.  For instance, we can establish whether a given theory admits certain charged branes. We also comment on the fact that quantum corrections can modify the supercurrent multiplet and show how these corrections are constrained by the structure of the multiplet.

Appendix A summarizes some facts about the energy-momentum tensor and its improvements.  Our conventions for two- and three-dimensional theories are summarized in appendix~B.  In appendix C, we describe the~$\CS$-multiplet in two-dimensional theories with~$\CN=(2,2)$ supersymmetry. Appendix D explains the relation between some additional supercurrents, which were discussed in~\refs{\GatesYC,\MagroAJ,\KuzenkoAM\ZhengXX\KuzenkoNI-\GatesNR}, and our general framework.

\newsec{Supercurrents in Four Dimensions}

In this section we show that the~$\CS$-multiplet~\fdsmult\ is the most general supercurrent satisfying the general requirements (a)--(d) laid out in the introduction. This multiplet must exist in any four-dimensional field theory with~$\CN=1$ supersymmetry. We then discuss the allowed improvements of the~$\CS$-multiplet and we use them to establish when the multiplet is decomposable. We will illustrate this using general Wess-Zumino models.

\subsec{Deriving the~$\CS$-Multiplet}

The most general supercurrent multiplet satisfying~(a)--(c) must contain a conserved supersymmetry current~$S_{\alpha\mu}$, a real, conserved, symmetric energy-momentum tensor~$T_{\mu\nu}$, and possibly other operators of lower spin. Since~$T_{\mu\nu}$ is the highest-spin operator, such a multiplet can be represented by a real superfield~$\CT_\mu$ with
\eqn\tprop{\CT_\mu \big|_{\theta\sigma^\nu \thetabar} \sim T_{\nu\mu} + \cdots~,}
where the ellipsis denotes lower-spin operators and their derivatives. The component structure of~$\CT_\mu$ must be consistent with the supersymmetry current algebra~\fdscurralgaa.  A detailed analysis shows that this completely fixes~$\CT_\mu$ and the Schwinger terms in~\fdscurralgaa. (We do not describe this arduous computation here.) Furthermore, the resulting expression for~$\CT_\mu$ is always decomposable. It can be separated into a sub-multiplet~$\CZ_\alpha$ and a smaller supercurrent\foot{The superfield~$\CZ_\alpha$ satisfies the defining relations
\eqn\Zsfcons{\eqalign{& D_\alpha \CZ_\beta + D_\beta \CZ_\alpha = 0~, \cr
& \b D^2 \CZ_\alpha + 2 \b D_\alphadot D_\alpha \b \CZ^\alphadot + D_\alpha \b D_\alphadot \b \CZ^\alphadot = 0~.}}
See appendix~D for a related discussion.}
\eqn\smultdef{\CS_{\alpha\alphadot} = \CT_{\alpha\alphadot} + i \left(D_\alpha \b \CZ_\alphadot + \b D_\alphadot \CZ_\alpha\right)~.}
This is the~$\CS$-multiplet~\fdsmult, which we repeat here for convenience:
\eqn\fdsmultrep{\eqalign{& \b D^\alphadot \CS_{\alpha\alphadot} = \chi_\alpha + \CY_\alpha~,\cr
& \b D_\alphadot \chi_\alpha = 0~,\qquad D^\alpha \chi_\alpha = \b D_\alphadot \b \chi^\alphadot~,\cr
& D_\alpha \CY_\beta + D_\beta \CY_\alpha = 0~,\qquad \b D^2 \CY_\alpha = 0~.}}
Thus, every supersymmetric field theory admits an~$\CS$-multiplet.

It is straightforward to solve the constraints~\fdsmultrep\ in components:
\eqn\fdsmultcomp{\eqalign{\CS_\mu  =~& j_\mu - i \theta \left(S_\mu - { i \over \sqrt2} \sigma_\mu \b \psi\right) + i \thetabar \left(\b S_\mu - { i \over \sqrt2} \b \sigma_\mu \psi\right) + {i \over 2} \thetasq \b Y_\mu - { i \over 2} \thetabarsq Y_\mu \cr
& + \left(\theta \sigma^\nu\thetabar\right) \left(2 T_{\nu\mu} - \eta_{\nu\mu} A - {1 \over 8} \ep_{\nu\mu\rho\sigma} F^{\rho\sigma} - \half \ep_{\nu\mu\rho\sigma} \d^\rho j^{\sigma} \right) \cr
& - \half \thetasq\thetabar\left(\b\sigma^\nu \d_\nu S_\mu +{i \over \sqrt2} \b \sigma_\mu \sigma^\nu \d_\nu \b \psi\right) + \half \thetabarsq \theta\left(\sigma^\nu \d_\nu \b S_\mu + {i \over \sqrt2} \sigma_\mu \b \sigma^\nu \d_\nu \psi\right) \cr
& + \half \thetasq \thetabarsq \left(\d_\mu \d^\nu j_\nu - \half \d^2 j_\mu \right)~.}}
The chiral superfield~$\chi_\alpha$ is given by
\eqn\fdchicomp{\eqalign{& \chi_\alpha  = - i \lambda_\alpha(y) + \theta_\beta \left({\delta_\alpha}^\beta D(y) - i {\left(\sigma^{\mu\nu}\right)_\alpha}^\beta F_{\mu\nu}(y)\right) + \thetasq \sigma^\mu_{\alpha\alphadot}\d_\mu\b\lambda^\alphadot(y)~,\cr
&   \lambda_\alpha = 2 \sigma^\mu_{\alpha\alphadot} {\b S^\alphadot}_\mu + 3 \sqrt2 i \psi_\alpha~,\cr
& D = -4 {T^\mu}_\mu + 6 A~,\cr
&  F_{\mu\nu} = -F_{\nu\mu}~, \qquad \d_{[\mu} F_{\nu\rho]} = 0~,}}
and the superfield~$\CY_\alpha$ is given by
\eqn\fdycomp{\eqalign{& \CY_\alpha  =  \sqrt2 \psi_\alpha + 2 \theta_\alpha F + 2i \sigma^\mu_{\alpha\alphadot}\thetabar^\alphadot Y_\mu - 2\sqrt2 i \left(\theta\sigma^\mu\thetabar\right) {\left(\sigma_{\mu\nu}\right)_\alpha}^\beta \d^\nu \psi_\beta \cr
& \hskip25pt + i \thetasq \sigma^\mu_{\alpha\alphadot} \thetabar^\alphadot \d_\mu F + \thetabarsq \theta_\alpha \d^\mu Y_\mu -{1 \over 2 \sqrt2} \thetasq \thetabarsq \d^2 \psi_\alpha~,\cr
&  \d_{[\mu} Y_{\nu]} = 0~, \cr
&  F = A + i \d^\mu j_\mu~.}}
The supersymmetry current~$S_{\alpha\mu}$ is conserved, and the energy-momentum tensor~$T_{\mu\nu}$ is real, conserved, and symmetric. The~$\CS$-multiplet contains~$16+16$ independent real operators.\foot{We define the number of independent operators as the number of components minus the number of conservation laws. For example, the $4 \times 5/2 = 10$ components of the energy-momentum tensor lead to $6$ independent operators, because there are~$4$ conservation laws.}

If there is a well-defined complex scalar~$x$ such that the complex closed one-form~$Y_\mu$ in~\fdycomp\ can be written as~$Y_\mu = \d_\mu x$, then we can express
\eqn\yx{\CY_\alpha = D_\alpha X~,\qquad \b D_\alphadot X = 0~,}
where the chiral superfield~$X$ is given by
\eqn\xsf{X = x(y) + \sqrt2 \theta \psi(y) + \thetasq F(y)~.}
In this case the~$\CS$-multiplet takes the form~\smultks\ discussed in~\KomargodskiRB. However, there are situations in which~$X$ does not exist and we must use~$\CY_\alpha$ (for an example, see subsection~2.3).

\subsec{Improvements and Decomposability}

The~$\CS$-multiplet is not unique. It can be modified by an improvement transformation,
\eqn\smultit{\eqalign{& \CS_{\alpha\alphadot} \rightarrow \CS_{\alpha\alphadot} + [D_\alpha, \b D_\alphadot] U~,\cr
& \chi_\alpha \rightarrow \chi_\alpha + {3 \over 2} \b D^2 D_\alpha U~,\cr
& \CY_\alpha \rightarrow \CY_\alpha + \half D_\alpha \b D^2 U~,}}
where the real superfield~$U$ takes the form
\eqn\ucomp{U = u + \theta \eta + \thetabar \b \eta + \thetasq N + \thetabarsq \b N - \left(\theta\sigma^\mu \thetabar\right) V_\mu + \cdots~.}
The transformation~\smultit\ preserves the constraints~\fdsmultrep. It modifies the supersymmetry current and the energy-momentum tensor by improvement terms as in~\scurrimp\ and~\symimp,
\eqn\STimp{\eqalign{& S_{\alpha\mu} \rightarrow S_{\alpha\mu} + 2  {\left(\sigma_{\mu\nu}\right)_\alpha}^\beta \d^\nu \eta_\beta~,\cr
& T_{\mu\nu} \rightarrow T_{\mu\nu} + \half\left(\d_\mu \d_\nu - \eta_{\mu\nu} \d^2\right)u~,}}
and it also shifts
\eqn\otherimps{\eqalign{
& F_{\mu\nu} \rightarrow F_{\mu\nu} - 6 \left(\d_\mu V_\nu - \d_\nu V_\mu\right)~,\cr
& Y_\mu \rightarrow Y_\mu - 2 \d_\mu \b N~.}}

In order for the improvement transformation~\smultit\ to be well-defined, the superfield~$U$ must be well-defined up to shifts by a real constant. It is possible to express this transformation entirely in terms of the well-defined superfield~$\zeta_\alpha = D_\alpha U$.\foot{The superfield~$\zeta_\alpha$ satisfies the constraints
\eqn\zetamult{\eqalign{& D_\alpha \zeta_\beta + D_\beta \zeta_\alpha = 0~,\cr
& \b D^2 \zeta_\alpha + 2 \b D_\alphadot D_\alpha \b \zeta^\alphadot + D_\alpha \b D_\alphadot \b \zeta^\alphadot = 0~.}}
In terms of~$\zeta_\alpha$, the improvement transformation~\smultit\ takes the form
\eqn\zetaimp{\eqalign{& \CS_{\alpha\alphadot} \rightarrow \CS_{\alpha\alphadot} + D_\alpha \b \zeta_\alphadot - \b D_\alphadot \zeta_\alpha~,\cr
& \chi_\alpha \rightarrow \chi_\alpha + {3 \over 2} \b D^2 \zeta_\alpha~,\cr
& \CY_\alpha \rightarrow \CY_\alpha + \half D_\alpha \b D_\alphadot \b \zeta^\alphadot~.}}
This is similar, but not identical, to the transformation~\smultdef, which involves the superfield~$\CZ_\alpha$ defined in \Zsfcons.} With this understanding and for ease of notation, we continue to work in terms of~$U$.

As we explained in the introduction, the~$\CS$-multiplet is reducible, since~$\chi_\alpha$ and~$\CY_\alpha$ are non-trivial sub-multiplets, but it is generally indecomposable. However, there are special cases in which we can use improvements~\smultit\ to decompose the~$\CS$-multiplet.  This gives rise to smaller supercurrent multiplets:
\medskip
\item{1.)} If there is a well-defined real~$U$ such that~$\chi_\alpha = -{3 \over 2} \b D^2 D_\alpha U$, then~$\chi_\alpha$ can be improved to zero. In this case the~$\CS$-multiplet decomposes into~$\chi_\alpha$ and a supercurrent~$\CJ_{\alpha\alphadot}$ satisfying
\eqn\FZmult{\eqalign{&\b D^\alphadot \CJ_{\alpha\alphadot} =\CY_\alpha~,\cr
& D_\alpha \CY_\beta + D_\beta \CY_\alpha = 0~,\qquad \b D^2 \CY_\alpha = 0~.}}
This is the~FZ-multiplet~\FerraraPZ. It contains~$12+12$ independent real operators. If it is possible to write~$\CY_\alpha = D_\alpha X$ as in~\yx, then we recover the more familiar form of the FZ-multiplet,
\eqn\famfzmult{\eqalign{&\b D^\alphadot \CJ_{\alpha\alphadot} =D_\alpha X~,\cr
& \b D_\alphadot  X = 0~.}}
\medskip
\item{2.)} If there is a well-defined real~$U$ such that~$X = - \half \b D^2 U$, then~$\CY_\alpha = D_\alpha X$ can be improved to zero. In this case the~$\CS$-multiplet decomposes into~$\CY_\alpha$ and a supercurrent~$\CR_{\alpha\alphadot}$ satisfying
\eqn\rmult{\eqalign{& \b D^\alphadot \CR_{\alpha\alphadot}  = \chi_\alpha~,\cr
&  \b D_\alphadot \chi_\alpha = 0~, \quad D^\alpha \chi_\alpha = \b D_\alphadot \b \chi^\alphadot~.}}
This is the~$\CR$-multiplet~\refs{\GatesYC,\DienesTD\KuzenkoYM-\KomargodskiRB,\GatesNR}. Like the FZ-multiplet, the~$\CR$-multiplet contains~$12+12$ independent real operators. The constraints~\rmult\ imply that~$\d^\mu \CR_\mu = 0$, so that the bottom component of~$\CR_\mu$ is a conserved~$R$-current. Conversely, any theory with a continuous~$R$-symmetry admits an~$\CR$-multiplet.
\medskip
\item{3.)} If we can set both~$\chi_\alpha$ and~$\CY_\alpha$ to zero by a single improvement transformation, then the theory is superconformal and the~$\CS$-multiplet decomposes into~$\chi_\alpha$, $\CY_\alpha$, and an~$8+8$ supercurrent~$\CJ_{\alpha\alphadot}$ satisfying
\eqn\scftmult{\b D^\alphadot \CJ_{\alpha\alphadot} = 0~.}

\bigskip

\noindent The FZ-multiplet and the~$\CR$-multiplet allow residual improvement transformations, which preserve the conditions~$\chi_\alpha = 0$ and~$\CY_\alpha = 0$ respectively.

With the exception of the special cases discussed above, the~$\CS$-multiplet is indecomposable. This is because we insist on discussing only well-defined operators.  Other authors have decomposed the~$\CS$-multiplet even when it is indecomposable, because they were willing to consider operators that are either not gauge invariant or not globally well-defined.

\subsec{The~$\CS$-Multiplet in Wess-Zumino Models}

As an example, we consider a general Wess-Zumino model with K\"ahler potential~$K(\Phi^i, \b \Phi^{\ibar})$ and superpotential~$W(\Phi^i)$. The K\"ahler potential and the superpotential need not be well-defined: $K$ may be shifted by K\"ahler transformations,
\eqn\kt{K(\Phi^i, \b \Phi^{\ibar}) \rightarrow K(\Phi^i, \b \Phi^{\ibar}) + \Lambda(\Phi^i) + \b \Lambda(\b \Phi^{\ibar})~,}
and~$W$ may be shifted by constants. This is because the component Lagrangian of the theory only depends on the K\"ahler metric~$g_{i \bar j} = \d_i \d_{\bar j} K$ of the target space, and on the derivatives~$\d_i W$ of the superpotential. Thus, only~$g_{i \jbar}$ and~$\d_i W$ must be well-defined. We can use the metric to construct the K\"ahler form
\eqn\kf{\Omega = i g_{i \jbar} d\Phi^i \wedge d\b\Phi^{\jbar}~,}
which is real and closed,~$d \Omega = 0$. Locally, it can be expressed as
\eqn\kconn{\Omega = d \CA~, \qquad \CA = - { i \over 2} \d_i K d \Phi^i + {i \over 2} \d_{\ibar} K d \b \Phi^{\ibar}~.} In general, the K\"ahler connection~$\CA$ is not globally well-defined.

Using the equations of motion~$\b D^2 \d_i K = 4 \d_i W$, we can check that the superfields
\eqn\wzsmult{\eqalign{& \CS_{\alpha\alphadot} = 2 g_{i \jbar} D_\alpha \Phi^i \b D_\alphadot \b \Phi^{\jbar}~,\cr
& \chi_\alpha = \b D^2 D_\alpha K~,\cr
& \CY_\alpha = 4 D_\alpha W~,}}
satisfy the constraints~\fdsmult. These operators are well-defined under K\"ahler transformations~\kt\ and shifts of~$W$ by a constant. Thus, the Wess-Zumino model has a well-defined~$\CS$-multiplet, as must be the case in any supersymmetric field theory. We would like to know under what conditions this multiplet is decomposable, so that it can be improved to an FZ-multiplet or an~$\CR$-multiplet.

If we take~$U = - {2 \over 3} K$ in~\smultit, then~$\chi_\alpha$ is improved to zero and we obtain an FZ-multiplet. This is allowed only if~$U \sim K$ is well-defined up to shifts by a real constant. In other words, the K\"ahler connection~$\CA$ must be globally well-defined~\KomargodskiRB. Note that this never happens on a compact manifold, where some power of~$\Omega$ is proportional to the volume form, which cannot be exact. As an example, consider a single chiral superfield~$\Phi$ with K\"ahler potential
\eqn\cpone{K = f^2 \log\left(1 + |\Phi|^2 \right)~,}
where~$f$ is a real constant of dimension one and~$\Phi$ is dimensionless. This K\"ahler potential gives rise to the Fubini-Study metric on~$\C\P^1$, which is compact. In this theory, the~$\CS$-multiplet cannot be improved to an FZ-multiplet.

If~$W$ is not well-defined, then it is not possible to express~$\CY_\alpha = D_\alpha X$ as in~\yx. Therefore, such a model cannot have an~$\CR$-multiplet. A simple example is a cylinder-valued chiral superfield~$\Phi \sim \Phi + 1$, with canonical K\"ahler potential and superpotential~$W \sim \Phi$. Going around the cylinder shifts~$W$ by a constant, and hence it is not well-defined.

If~$W$ is well-defined, then so is~$X = 4W$. We can improve~$X$ to zero and obtain an~$\CR$-multiplet if and only if the theory has a continuous~$R$-symmetry.  This requires a basis in which the fields~$\Phi^i$ can be assigned~$R$-charges~$R_i$ such that the superpotential has~$R$-charge~$2$,
\eqn\wid{2 W = \sum_i R_i \Phi^i \d_i W~,}
and the K\"ahler potential is~$R$-invariant up to a K\"ahler transformation,
\eqn\kid{\sum_i \left( R_i \Phi^i \d_i K - R_i \b \Phi^{\ibar} \d_{\ibar} K \right) = \Xi(\Phi^j) + \b \Xi(\b \Phi^{\jbar})~.}
Using~\wid\ and the equations of motion, we can write
\eqn\widii{X = - \half \b D^2U~, \qquad U = -  \sum_i R_i \Phi^i \d_i K~.}
This~$U$ is real as long as the chiral superfield~$\Xi$ in~\kid\ is a constant. In other words, $K$ must be~$R$-invariant up to shifts by a real constant. Furthermore, $U$ is well-defined as long as we only perform K\"ahler transformations that preserve this~$R$-invariance of~$K$. If both of these conditions are satisfied, then we can use~$U$ in~\smultit\ to obtain an~$\CR$-multiplet.

For example, the~$\C\P^1$ model~\cpone\ has an~$\CR$-multiplet. However, the cylinder-valued superfield~$\Phi \sim \Phi + 1$ with canonical~$K$ and~$W \sim \Phi$ does not have an~$\CR$-multiplet. This follows from the fact that the theory does not have a well-defined~$X$.  More explicitly, the superpotential~$W \sim \Phi$ forces us to assign~$R_\Phi = 2$, so that the $R$-transformation multiplies the bottom component of~$\Phi$ by a phase, but this is incompatible with the cylindrical field space $\Phi \sim \Phi+1$.

\newsec{Physical Interpretation in Terms of Brane Currents}

We have seen that the~$\CS$-multiplet, though generally indecomposable, can sometimes be improved to a smaller supercurrent multiplet. This is possible whenever~$\chi_\alpha$ or~$\CY_\alpha$ can be expressed in terms of a real superfield~$U$. The non-existence of such a~$U$ is an obstruction to the decomposability of the~$\CS$-multiplet. In this section we interpret this obstruction physically.

Let us consider the current algebra that follows from the~$\CS$-multiplet,\foot{We use the fact that~$[\xi^\alpha Q_\alpha + \b \xi_\alphadot \b Q^\alphadot, S] = i ( \xi^\alpha \CQ_\alpha + \b \xi_\alphadot \b \CQ^\alphadot) S~$,
for any superfield~$S$. Here~$Q_\alpha$ is the supercharge and~$\CQ_\alpha$ is the corresponding superspace differential operator. The additional factor of~$i$ is needed for consistency with Hermitian conjugation.}
\eqn\fdscurralg{\eqalign{&\{ \b Q_\alphadot, S_{\alpha \mu}\} = \sigma^\nu_{\alpha\alphadot} \left( 2 T_{\nu\mu} - {1 \over 8}\ep_{\nu\mu\rho\sigma} F^{\rho\sigma}  + i \d_\nu j_\mu - i \eta_{\nu\mu} \d^\rho j_\rho- \half \ep_{\nu\mu\rho\sigma} \d^\rho j^{\sigma}\right)~,\cr
& \{Q_\beta, S_{\alpha \mu}\} = 2i \left( \sigma_{\mu\nu}\right)_{\alpha\beta} \b Y^\nu~.}}
Recall from~\fdchicomp\ and~\fdycomp\ that the real closed two-form~$F_{\mu\nu}$ is embedded in~$\chi_\alpha$ and that the complex closed one-form~$Y_\mu$ is embedded in~$\CY_\alpha$. To elucidate the role of these operators, we define
\eqn\branecurr{\eqalign{& C_{\mu\nu} = - {1 \over 16} \ep_{\mu\nu\rho\sigma} F^{\rho\sigma}~,\qquad \d^\nu C_{\mu\nu} = 0~,\cr
& C_{\mu\nu\rho} = - \ep_{\mu\nu\rho\sigma} \b Y^\sigma~,\qquad \d^\rho C_{\mu\nu\rho} = 0~.}}
The current algebra~\fdscurralg\ then takes the form~\fdscurralgaa. We see that the Schwinger terms depend only on~$j_\mu$ and that there are no such terms in~$\{Q_\beta, S_{\alpha\mu}\}$. As we mentioned in the introduction, the two-form current~$C_{\mu\nu}$ is associated with strings and the three-form current~$C_{\mu\nu\rho}$ is associated with domain walls. The appearance of such currents in the four-dimensional~$\CN=1$ current algebra was pointed out in~\refs{\DvaliXE,\GorskyHK}.

We can formally define the string and domain wall charges
\eqn\sdwch{Z_\mu = \int d^3 x\, {C_\mu}^0~,\qquad Z_{\mu\nu} = \int d^3 x \, {C_{\mu\nu}}^0~,}
and integrate the current algebra~\fdscurralg\ to obtain the modified supersymmetry algebra~\extfdsalg, which we repeat here for convenience:
\eqn\extsalg{\eqalign{&\{Q_\alpha, \b Q_\alphadot\} = 2 \sigma^\mu_{\alpha\alphadot} \left( P_\mu + Z_\mu\right)~,\cr
& \{Q_\alpha, Q_\beta\} = \sigma^{\mu\nu}_{\alpha\beta} Z_{\mu\nu}~.}}
In obtaining this algebra, we have dropped the contributions from the~$j_\mu$-dependent Schwinger terms in~\fdscurralg, which can contribute a boundary term to~$\{Q_\alpha, \b Q_\alphadot\}$. The imaginary part of this boundary term must vanish by unitarity. The real part is due to the term~$\sim \ep_{\mu\nu\rho\sigma} \d^\rho j^\sigma$ in~\fdscurralg, and we assume that it vanishes as well. (We will revisit this point below.) Note that the string charge~$Z_\mu$ is algebraically indistinguishable from the momentum~$P_\mu$. However, they are distinguished at the level of the current algebra~\fdscurralg.

As we mentioned in the introduction, the brane charges~$Z_\mu$ and~$Z_{\mu\nu}$ are not central charges of the super-Poincar\'e algebra. Moreover, they are generally infinite, and only the charge per unit volume is meaningful. For instance, it determines the tension of BPS branes. Many authors have studied such BPS configurations (see e.g.~\refs{\GauntlettCH,\ShifmanZZ} and references therein). Our new point here is the relation between the brane currents and the different supercurrent multiplets.

Under improvements~\smultit\ of the~$\CS$-multiplet, the shifts of~$F_{\mu\nu}$ and~$Y_\mu$ in~\otherimps\ imply that the brane currents also change by improvement terms,
\eqn\braneimp{\eqalign{& C_{\mu\nu} \rightarrow C_{\mu\nu} + {3 \over 4} \ep_{\mu\nu\rho\sigma} \d^\rho V^\sigma~,\cr
& C_{\mu\nu\rho} \rightarrow C_{\mu\nu\rho} + 2 \ep_{\mu\nu\rho\sigma} \d^\sigma N~,}}
where~$V_\mu$ and~$N$ belong to the superfield~$U$ in~\ucomp. Upon integration, these improvement terms can contribute boundary terms to the brane charges~$Z_\mu$ and~$Z_{\mu\nu}$. Whether or not such boundary terms arise depends on the behavior of~$V_\mu$ and~$N$ at spatial infinity.  Note that the Schwinger term~$\sim \ep_{\mu\nu\rho\sigma} \d^\rho j^\sigma$ in~\fdscurralg\ looks like an improvement term for~$C_{\mu\nu}$ with~$V_\mu \sim j_\mu$. As long as~$j_\mu$, $V_\mu$, and~$N$ are sufficiently well-behaved at spatial infinity, all boundary terms vanish and the brane charges are not affected by the improvements~\braneimp. This is the case for isolated branes, as long as the fields approach a supersymmetric vacuum far away from the brane.\foot{In the presence of more complicated configurations, such as certain brane bound states, this is no longer true. For instance, the Schwinger term~$\sim \ep_{\mu\nu\rho\sigma} \d^\rho j^\sigma$ in~\fdscurralg\ gives rise to a boundary contribution in the presence of domain wall junctions~\refs{\GorskyHK,\RitzMP}. However, the string-like defect on which the domain walls end does not exist in isolation, and hence this boundary term is not a conventional string charge.} The fact that improvements of the supersymmetry current do not affect the brane charges was pointed out in~\refs{\ChibisovRC,\GorskyHK}.

With these assumptions, we conclude that the string charge~$Z_\mu$ must vanish in theories in which~$F_{\mu\nu}$ can be set to zero by an improvement transformation. This is the case if and only if the~$\CS$-multiplet can be improved to an FZ-multiplet. Likewise, the domain wall charge~$Z_{\mu\nu}$ must vanish in theories in which~$Y_\mu$ can be set to zero by an improvement transformation, and this happens if and only if the~$\CS$-multiplet can be improved to an~$\CR$-multiplet. Conversely, the existence of strings that carry charge~$Z_\mu$ is a physical obstruction to improving the~$\CS$-multiplet to an FZ-multiplet, and the existence of domain walls that carry charge~$Z_{\mu\nu}$ is a physical obstruction to improving the~$\CS$-multiplet to an~$\CR$-multiplet.\foot{This does not apply to branes whose charges do not appear in the supersymmetry algebra. For instance, there can be strings in theories with FZ-multiplets, provided the string charge does not appear in the supersymmetry algebra~\VainshteinHU. Clearly, such strings cannot be BPS.}

This point of view emphasizes the fact that the~$\CS$-multiplet always exists, but that it may be decomposable. The existence of brane charges in the supersymmetry algebra is an obstruction to decomposability, and it forces us to consider different supercurrents containing the corresponding brane currents: charged domain walls lead to the FZ-multiplet, and charged strings give rise to the~$\CR$-multiplet. Theories that support both domain walls and strings with charges in the supersymmetry algebra require the~$\CS$-multiplet.

To illustrate this, we return to the Wess-Zumino models of subsection~2.3. If such a model admits strings with charge~$Z_\mu$, then~$\chi_\alpha = \b D^2 D_\alpha K$ in~\wzsmult\ cannot be improved to zero. Therefore, the K\"ahler form~$\Omega$ in~\kt\ is not exact. In this case, the operator~$F_{\mu\nu}$ in the~$\CS$-multiplet is proportional to the pull-back to spacetime of~$\Omega$, and the string current~$C_{\mu\nu} \sim i \ep_{\mu\nu\rho\sigma} g_{i\jbar}\d^\rho\phi^i \d^\sigma \b \phi^{\jbar} $ is topological. (This is familiar in the context of two-dimensional sigma models, where the analogues of four-dimensional strings are instantons.) If the string is oriented along the~$z$-axis in its rest frame, then the string charge is given by~$Z_\mu = \pm T_{\rm BPS} L \delta_{\mu 3}$, where~$T_{\rm BPS} >0$ is a constant,~$L \rightarrow \infty$ is the length of the string, and the sign is determined by the chirality of the string. From~\extsalg, we see that the mass~$M$ of the string satisfies the BPS bound~$M \geq T_{\rm BPS} L$. If this bound is saturated, then the string has tension~$T_{\rm BPS}$, and it preserves two real supercharges. A typical example is the~$\C\P^1$ model~\cpone, which supports BPS strings with~$T_{\rm BPS} \sim f^2$.

If the Wess-Zumino model admits domain walls that carry charge~$Z_{\mu\nu}$, then~$\CY_\alpha$ cannot be improved to zero. Hence, the theory does not have a continuous~$R$-symmetry. In this case, the operator~$Y_\mu$ in the~$\CS$-multiplet is proportional to the pull-back to spacetime of the holomorphic one-form~$\d_i W d\Phi^i$. If the domain wall is at rest and lies in the~$xy$-plane, then the non-vanishing components of the domain wall charge are~$Z_{12} = -Z_{21} = 2 z_{\rm BPS} A$, where~$z_{\rm BPS}$ is a complex constant and~$A\rightarrow \infty$ is the area of the wall. From~\extsalg, we see that the mass~$M$ of the wall satisfies the BPS bound~$M \geq |z_{\rm BPS}| A$. If this bound is saturated, then the wall has tension~$|z_{\rm BPS}|$, and it preserves two real supercharges. A simple example is a single chiral superfield~$\Phi$ with canonical K\"ahler potential and superpotential~$W = {m \over 2} \Phi^2 + {\lambda \over 3} \Phi^3$. This model has two degenerate supersymmetric vacua, and it supports a BPS domain wall, which interpolates between them. In this case~$z_{\rm BPS} = -2 \Delta \b W$, where~$\Delta W = \pm {m^3 \over 6 \lambda^2}$ is the difference of the superpotential evaluated in the two vacua; the sign is determined by the choice of vacuum on either side of the wall.

\newsec{Supercurrents in Three Dimensions}

In this section, we discuss the analogue of the~$\CS$-multiplet in three-dimensional theories with~$\CN=2$ supersymmetry. (Our conventions are summarized in appendix~B.) Just as in four-dimensions, this multiplet is the most general supercurrent satisfying the requirements~(a)--(d) laid out in the introduction. Consequently, it exists in every supersymmetric field theory.

\subsec{The~$\CS$-Multiplet}

In three-dimensional~$\CN=2$ theories, the~$\CS$-multiplet is a real superfield~$\CS_\mu$, which  satisfies the constraints
\eqn\tdsmult{\eqalign{& \b D^\beta \CS_{\alpha\beta} = \chi_\alpha + \CY_\alpha~,\cr
& \b D_\alpha \chi_\beta = \half C \ep_{\alpha\beta}~, \qquad D^\alpha\chi_\alpha = -\b D^\alpha \b \chi_\alpha~,\cr
& D_\alpha\CY_\beta + D_\beta\CY_\alpha = 0~, \qquad \b D^\alpha \CY_\alpha = -C~,}}
where~$\CS_{\alpha\beta} = \CS_{\beta\alpha}$ is the symmetric bi-spinor corresponding to~$\CS_\mu$, and~$C$ is a complex constant. We will see that~$C$ gives rise to a new kind of brane current, which is qualitatively different from the brane currents we encountered in four dimensions.

It is straightforward to solve the constraints~\tdsmult\ in components:
\eqn\tdsmultcomp{\eqalign{\eqalign{ \CS_\mu  =~& j_\mu - i \theta \left( S_\mu + {i \over \sqrt2} \gamma_\mu \b \psi\right) - i \thetabar \left( \b S_\mu - {i \over \sqrt2} \gamma_\mu \psi\right) + {i \over 2} \thetasq \b Y_\mu + {i \over 2} \thetabarsq Y_\mu\cr
& - \left(\theta \gamma^\nu \thetabar\right) \left( 2 T_{\nu\mu} -  \eta_{\mu\nu} A + { 1\over 4} \ep_{\nu\mu\rho}H^\rho\right) - i \theta\thetabar \left( {1 \over 4} \ep_{\mu\nu\rho} F^{\nu\rho} +  \ep_{\mu\nu\rho} \d^\nu j^\rho\right) \cr
&+ \half \thetasq \thetabar\left( \gamma^\nu \d_\nu S_\mu - {i \over \sqrt2} \gamma_\mu \gamma_\nu \d^\nu \bar \psi\right) + \half \thetabarsq \theta \left( \gamma^\nu \d_\nu \b S_\mu + {i \over \sqrt2} \gamma_\mu \gamma_\nu \d^\nu \psi\right) \cr
&-\half \thetasq \thetabarsq \left( \d_\mu\d^\nu j_\nu - \half \d^2 j_\mu\right)~.}}}
The chiral superfield~$\chi_\alpha$ is given by
\eqn\tdchicomp{\eqalign{& \chi_\alpha  =  - i \lambda_\alpha(y) +  \theta_\beta \Big({\delta_\alpha}^\beta D(y) - {{\gamma^\mu}_\alpha}^\beta \big( H_\mu(y) - { i \over 2} \ep_{\mu\nu\rho} F^{\nu\rho}(y)\big)\Big) \cr
& \hskip24pt + \half  \thetabar_\alpha C - \thetasq {{\gamma^\mu}_\alpha}^\beta \d_\mu\b\lambda_\beta(y)~,\cr
& \lambda_\alpha = -2 {{\gamma^\mu}_\alpha}^\beta {\b S}_{\beta\mu} + 2 \sqrt2 i \psi_\alpha~, \cr
& D = -4 {T^\mu}_\mu + 4 A~,\cr
& \d_{[\mu} H_{\nu]}  =0~, \cr
&  F_{\mu\nu} = -F_{\nu\mu}~, \qquad \d_{[\mu} F_{\nu\rho]} = 0~,\cr
&y^\mu = x^\mu - i \theta \gamma^\mu \thetabar~,}}
and the superfield~$\CY_\alpha$ is given by 
\eqn\tdycomp{\eqalign{& \CY_\alpha = \sqrt2 \psi_\alpha + 2 \theta_\alpha F - \half  \thetabar_\alpha C + 2i \gamma^\mu_{\alpha\beta} \thetabar^\beta Y_\mu + \sqrt 2 i\left(\theta\gamma^\mu\thetabar\right) \ep_{\mu\nu\rho} {{\gamma^\nu}_\alpha}^\beta \d^\rho \psi_\beta \cr
& \hskip25pt + \sqrt2 i \theta\thetabar {{\gamma^\mu}_\alpha}^\beta \d_\mu \psi_\beta + i \thetasq \gamma^\mu_{\alpha\beta} \thetabar^\beta \d_\mu F - \thetabarsq \theta_\alpha \d^\mu Y_\mu + {1 \over 2 \sqrt2} \thetasq \thetabarsq \d^2 \psi_\alpha~,\cr
& \d_{[\mu} Y_{\nu]} = 0~, \cr
& F = A + i \d^\mu j_\mu~.}}
The supersymmetry current~$S_{\alpha\mu}$ is conserved, and the energy-momentum tensor~$T_{\mu\nu}$ is real, conserved, and symmetric. The~$\CS$-multiplet now contains~$12+12$ independent real operators, and the complex constant~$C$.

If there is a well-defined complex scalar~$x$ such that the complex closed one-form~$Y_\mu$ in~\tdycomp\ can be written as~$Y_\mu = \d_\mu x$, then we can express
\eqn\chixtd{\CY_\alpha = D_\alpha X~, \qquad D_\alpha \b D_\beta X = - \half C\ep_{\alpha\beta}~, \qquad \b D^2 X = 0~,}
where~$X| = x$. If the constant~$C$ vanishes, then~$X$ is chiral, just as in four dimensions.

If there is a well-defined real scalar~$J$ such that the real closed one-form~$H_\mu$ in~\tdchicomp\ can be written as~$H_\mu = \d_\mu J$, then we can express
\eqn\chijtd{\chi_\alpha = i \b D_\alpha \CJ~, \qquad \b D^2 \CJ = -i C~,}
where~$\CJ| = J$. If the constant~$C$ vanishes, then~$\CJ$ is a real linear multiplet.\foot{A real linear multiplet~$\CO$ satisfies~$\b D^2 \CO = 0$ and hence also~$D^2 \CO = 0\,$.}

\subsec{Improvements and Decomposability}

The~$\CS$-multiplet~\tdsmult\ can be modified by an improvement transformation
\eqn\tdsmulit{\eqalign{& \CS_{\alpha\beta} \rightarrow \CS_{\alpha\beta} + \half \left([D_\alpha, \b D_\beta] + [D_\beta, \b D_\alpha]\right) U ~,\cr
& \chi_\alpha \rightarrow \chi_\alpha - \b D^2 D_\alpha U~,\cr
& \CY_\alpha \rightarrow \CY_\alpha - \half D_\alpha \b D^2 U~,}}
where the real superfield~$U$ takes the form
\eqn\tducomp{U = \cdots + \thetasq N - \thetabarsq \b N + \left(\theta\gamma^\mu\thetabar\right) V_\mu - i \theta\thetabar K + \cdots~.}
The transformation~\tdsmulit\ preserves the constraints~\tdsmult, and it changes the supersymmetry current and the energy-momentum tensor by improvement terms. It also shifts
\eqn\tdbcshift{\eqalign{& H_\mu \rightarrow  H_\mu - 4 \d_\mu K~,\cr
& F_{\mu\nu} \rightarrow F_{\mu\nu} - 4 \left(\d_\mu V_\nu - \d_\nu V_\mu\right)~,\cr
& Y_\mu \rightarrow Y_\mu -2 \d_\mu \b N~.}}
The constant~$C$ is not affected. As in four dimensions, the superfield~$U$ in~\tdsmulit\ is only well-defined up to shifts by a real constant, and we could instead work with the well-defined superfield~$\zeta_\alpha = D_\alpha U$.\foot{In three dimensions, the superfield~$\zeta_\alpha$ satisfies the constraints (compare with~\zetamult)
\eqn\tdzetacons{\eqalign{& \b D^\alpha \zeta_\alpha = D^\alpha \b \zeta_\alpha~,\cr
& D_\alpha \zeta_\beta + D_\beta \zeta_\alpha = 0~,\cr
& \b D^2 \zeta_\alpha + 2 \b D^\beta D_\alpha \b \zeta_\beta + D_\alpha \b D^\beta \b \zeta_\beta = 0~.}}}

Again, we distinguish cases in which the~$\CS$-multiplet can be improved to a smaller supercurrent:
\smallskip
\item{1.)} If~$C = 0$ and there is a well-defined real~$U$ such that~$\CJ = 2 i \b D D U$, then~$\chi_\alpha =  i \b D_\alpha \CJ$ can be improved to zero and we obtain an FZ-multiplet
    \eqn\tdfz{\eqalign{&\b D^\beta \CJ_{\alpha\beta} = \CY_\alpha~,\cr
    & D_\alpha \CY_\beta + D_\beta \CY_\alpha = 0~,\qquad \b D^\alpha \CY_\alpha = 0~.}}
    This multiplet contains~$8+8$ independent real operators.
\medskip
\item{2.)} If~$C = 0$ and there is a well-defined real~$U$ such that~$X = \half \b D^2 U$, then~$\CY_\alpha = D_\alpha X$ can be improved to zero and we obtain an~$\CR$-multiplet
    \eqn\tdrmult{\eqalign{& \b D^\beta \CR_{\alpha\beta} = \chi_\alpha~,\cr
    & \b D_\alpha \chi_\beta = 0~, \qquad D^\alpha \chi_\alpha = - \b D^\alpha \b \chi_\alpha~.}}
    Like the FZ-multiplet, it contains~$8+8$ independent real operators. As in four dimensions, the bottom component of the~$\CR$-multiplet is a conserved~$R$-current, and the~$\CR$-multiplet exists in every theory with a continuous~$R$-symmetry.
\medskip
\item{3.)} If~$C$ = 0 and we can set both~$\chi_\alpha$ and~$\CY_\alpha$ to zero by a single improvement transformation, then the theory is superconformal, and it has a~$4+4$ multiplet satisfying
    \eqn\tdscft{\b D^\beta \CJ_{\alpha\beta} = 0~.}
\medskip
\noindent Note that the~$\CS$-multiplet is decomposable only if the constant~$C$ vanishes.

\subsec{Brane Currents}

The current algebra that follows from the~$\CS$-multiplet takes the form
\eqn\tdsusyvar{\eqalign{& \{ \b Q_\alpha, S_{\beta \mu} \} = \gamma^\nu_{\alpha\beta} \left( 2 T_{\nu\mu} + {1 \over 4} \ep_{\nu\mu\rho} H^\rho + i \d_\nu j_\mu - i \eta_{\mu\nu} \d^\rho j_\rho \right) \cr
 & \hskip 55pt + i \ep_{\alpha\beta} \left( {1 \over 4} \ep_{\mu\nu\rho}F^{\nu\rho} + \ep_{\mu\nu\rho} \d^\nu j^\rho\right)~,\cr
& \{Q_\alpha, S_{\beta\mu}\} = {1 \over 4} \b C \gamma_{\mu\alpha\beta} + i \ep_{\mu\nu\rho} \gamma^\nu_{\alpha\beta} \b Y^\rho~.}}
This allows us to identify the conserved brane currents
\eqn\tdbranecurr{C_\mu\sim  \ep_{\mu\nu\rho} F^{\nu\rho}~,\qquad C_{\mu\nu} \sim  \ep_{\mu\nu\rho} H^\rho~,\qquad C'_{\mu\nu} \sim  \ep_{\mu\nu\rho} \b Y^\rho~,\qquad C_{\mu\nu\rho} \sim \b C \ep_{\mu\nu\rho}~.}
The current~$C_\mu$ is associated with zero-branes (it gives rise to a well-defined central charge~$Z$;~see appendix~B), while~$C_{\mu\nu}$ and~$C'_{\mu\nu}$ are associated with one-branes. The current~$C_{\mu\nu\rho}$ is associated with space-filling two-branes.

As in four-dimensions, improvement transformations~\tdsmulit\ of the~$\CS$-multiplet shift the brane-currents~\tdbranecurr\ by improvement terms, so that the corresponding brane charges are unchanged. (The space-filling brane current~$C_{\mu\nu\rho}$ is not affected.) Thus, the zero-brane charge corresponding to~$C_\mu$ and the one-brane charge corresponding to~$C_{\mu\nu}$ must vanish, if the~$\CS$-multiplet can be improved to an FZ-multiplet.  The one-brane charge corresponding to~$C'_{\mu\nu}$ must vanish, if the~$\CS$-multiplet can be improved to an~$\CR$-multiplet. Conversely, the existence of branes carrying these charges is a physical obstruction to improving the~$\CS$-multiplet to one of the smaller supercurrents.

\subsec{Relation to Four-Dimensional Supercurrents}

It is instructive to reduce the four-dimensional~$\CS$-multiplet~\fdsmult\ to three dimensions. Upon reduction, the four-dimensional superfield~$\CS_{\alpha\betadot}$ decomposes into a symmetric bi-spinor~$\hat \CS_{\alpha\beta}$ and a real scalar~$\hat \CJ$, which arises as the component of the four-dimensional~$\CS_\mu$ in the reduced direction. Thus~$\hat\CJ$ contains a conserved current corresponding to translations in the reduced direction. The four-dimensional superfields~$\chi_\alpha, \CY_\alpha$ reduce to~$\hat \chi_\alpha, \hat \CY_\alpha$. The constraints~\fdsmult\ then take the form
\eqn\tdred{\eqalign{& \b D^\beta \hat\CS_{\alpha\beta} = 2 i \b D_\alpha \hat\CJ + \hat \chi_\alpha + \hat \CY_\alpha~,\cr
& \b D_\alpha \hat \chi_\beta = 0~, \qquad D^\alpha \hat \chi_\alpha = - \b D^\alpha \b{\hat  \chi}_\alpha~, \cr
& D_\alpha \hat \CY_\beta + D_\beta \hat \CY_\alpha = 0~, \qquad \b D^2  \hat \CY_\alpha = 0~.}}
These constraints imply that~$\b D^\alpha \hat \CY_\alpha = -C$, where~$C$ is a complex constant, and thus
\eqn\curconseq{\b D^2 \hat \CJ = - {i C \over 2}~.}
The constant~$C$ arises from the four-dimensional domain wall current~$C_{\mu\nu\rho}$ in~\branecurr, but in three dimensions it represents a space-filling brane current. We identify~\tdred\ as a three-dimensional~$\CS$-multiplet~\tdsmult\ with
\eqn\tdopid{\chi_\alpha = \hat\chi_\alpha + 2 i \b D_\alpha \hat \CJ~.}
In general, $\hat \CJ$ is non-trivial, so that it cannot be set to zero by a three-dimensional improvement transformation~\tdsmulit.

We see that the four-dimensional~$\CS$-multiplet, which has~$16+16$ independent operators, becomes decomposable upon reduction to three dimensions. It decomposes into a three-dimensional~$\CS$-multiplet, which has~$12+12$ independent operators, and another~$4+4$ multiplet. Likewise, the reduction of the four-dimensional~$\CR$-multiplet~\rmult\ decomposes into a three-dimensional~$\CR$-multiplet~\tdrmult, and another~$4+4$ multiplet. This is expected, because a continuous~$R$-symmetry is preserved by dimensional reduction.

However, the four-dimensional FZ-multiplet~\FZmult, which has~$12+12$ independent operators, reduces to a three-dimensional~$\CS$-multiplet~\tdsmult, which is generally indecomposable. This is because~$\hat \CJ$ gives rise to a non-trivial~$\chi_\alpha$ in~\tdopid, even when~$\hat \chi_\alpha = 0$.

\newsec{Supercurrents in Two Dimensions}

In this section, we present the analogue of the~$\CS$-multiplet in two-dimensional theories with~$\CN = (0,2)$ supersymmetry. (Our conventions are summarized in appendix~B.) In appendix~C we extend our results to theories with~$\CN=(2,2)$ supersymmetry.

In two-dimensional~$\CN = (0,2)$ theories, the~$\CS$-multiplet consists of two real superfields~$\CS_{++}$, $\CT_{----}$ and a complex superfield~$\CW_-$, which satisfy the constraints
\eqn\hetsmult{\eqalign{& \d_{--} \CS_{++} = D_+\CW_- - \b D_+ \b \CW_-~,\cr
& \b D_+ \CT_{----} = \half \d_{--} \CW_-~,\cr
& \b D_+ \CW_- = C~.}}
Here~$C$ is a complex constant. As in three dimensions, it is associated with a space-filling brane current.

It is straightforward to solve the constraints~\hetsmult\ in components:
\eqn\hetsmultcomp{\eqalign{& \CS_{++} = j_{++} - i \theta^+ S_{+++} - i \thetabar^+ \b S_{+++} - \theta^+\thetabar^+ T_{++++}~,\cr
& \CW_- = - \b S_{+--} - i \theta^+ \left(T_{++--} + {i \over 2} \d_{--} j_{++} \right) - \thetabar^+ C + {i \over 2} \theta^+\thetabar^+ \d_{++} \b S_{+--}~,\cr
& \CT_{----} = T_{----} - \half \theta^+ \d_{--} S_{+--} + \half \thetabar^+ \d_{--} \b S_{+--} + {1 \over 4} \theta^+ \thetabar^+ \d_{--}^2 j_{++}~.}}
The supersymmetry current is conserved, and the energy-momentum tensor is real, conserved, and symmetric,
\eqn\twodconseq{\eqalign{& \d_{++} S_{+--} + \d_{--} S_{+++} = 0~,\cr
& \d_{++} T_{\pm\pm--} + \d_{--} T_{\pm\pm++} = 0~,\cr
& T_{++--} = T_{--++}~.}}
Thus, the~$\CS$-multiplet contains~$2+2$ independent real operators,\foot{We count independent operators according to the rules explained in footnote~3. This can obscure the counting in two dimensions. For instance, we do not count left-moving operators, which satisfy~$\d_{++}\CO=0$.} and the constant~$C$. Note that~$j_{++}$ is not in general accompanied by another real operator~$j_{--}$.

The improvements of the~$\CS$-multiplet~\hetsmult\ take the form
\eqn\hetimp{\eqalign{& \CS_{++} \rightarrow \CS_{++} + [D_+, \b D_+] U~,\cr
& \CW_- \rightarrow \CW_- + \d_{--} \b D_+ U~,\cr
& \CT_{----} \rightarrow \CT_{----} + \half \d_{--}^2 U~,}}
where~$U$ is a real superfield, whose bottom component is well-defined up to shifts by a real constant. The transformation~\hetimp\ preserves the constraints~\hetsmult\ and it changes the supersymmetry current and the energy-momentum tensor by improvement terms. The constant~$C$ is not affected.

As before, we distinguish special cases:
\smallskip
\item{1.)} If~$C=0$ and there is a well-defined real superfield~$\CR_{--}$ such that~$\CW_- = i \b D_+ \CR_{--}$, we obtain an~$\CR$-multiplet
    \eqn\hetrmult{\eqalign{& \d_{--} \CR_{++} + \d_{++} \CR_{--} = 0~,\cr
    & \b D_+ \left( \CT_{----} - {i \over 2} \d_{--} \CR_{--} \right) = 0~.}}
    Here we have relabeled~$\CS_{++} \rightarrow \CR_{++}$. The bottom components of~$\CR_{\pm\pm}$ form a conserved~$R$-current with~$R = -{1 \over 4} \int dx \left(j_{++} + j_{--}\right)$.  Unlike in higher dimensions, the~$\CR$-multiplet now includes the same number~$(2+2)$ of independent real operators as  the~$\CS$-multiplet: the conserved, symmetric energy-momentum tensor, the conserved~$R$-current, and two conserved supersymmetry currents.
\medskip
\item{2.)} If~$C=0$ and we can set~$\CW_-$ to zero by an improvement transformation, the theory is superconformal and the~$\CS$-multiplet decomposes into the right-moving supercurrent
    \eqn\hetscft{\d_{--} \CS_{++} = 0~,}
    and the left-moving component~$T_{----}$ of the energy-momentum tensor.

\bigskip

The current algebra that follows from the~$\CS$-multiplet takes the from
\eqn\hetcurralg{\eqalign{& \{ \b Q_+, S_{+++}\} = -T_{++++} - {i \over 2} \d_{++} j_{++}~,\cr
& \{ \b Q_+, S_{+--} \} = - T_{++--} + {i \over 2} \d_{--} j_{++}~,\cr
& \{Q_+, S_{+++}\} = 0~,\cr
& \{Q_+, S_{+--} \} = i \b C~.}}
As in three dimensions, we interpret the constant~$C$ as a space-filling brane current. This brane current is not affected by improvement transformations~\hetimp, and it must vanish whenever the theory admits an~$\CR$-multiplet~\hetrmult.

\newsec{Examples}

\subsec{Fayet-Iliopoulos Terms}

Consider a free~$U(1)$ gauge theory with an FI-term in four dimensions:
\eqn\FIlag{{\scr L} = {1 \over 4 e^2} \int d^2 \theta \, W^\alpha W_\alpha + {\rm c.c.} + \xi \int d^4 \theta \, V~.}
Here~$W_\alpha = -{1 \over 4} \b D^2 D_\alpha V$ is the usual field-strength superfield. Using the equations of motion~$D^\alpha W_\alpha = e^2 \xi$, we find that this theory has an~$\CR$-multiplet
\eqn\firmult{\eqalign{& \CR_{\alpha\alphadot} = - {4 \over e^2} W_\alpha \b W_\alphadot~,\cr
& \chi_\alpha = -4 \xi W_\alpha~.}}
It cannot be improved to an FZ-multiplet. Such an improvement would require~$U \sim \xi V$ in~\smultit, and this is not gauge invariant~\refs{\KomargodskiPC,\KomargodskiRB}. If we couple~\FIlag\ to matter with a generic superpotential, there will no longer be a continuous~$R$-symmetry. In this case the theory has an indecomposable~$\CS$-multiplet; it admits neither an~$\CR$-multiplet nor an FZ-multiplet.

We see that~$\chi_\alpha \sim \xi W_\alpha$ cannot be improved to zero in theories with an FI-term, and therefore they do not have an FZ-multiplet. From our discussion in section~3 we expect these theories to admit strings carrying charge~$Z_\mu$. This is indeed the case: even the simplest nontrivial example, supersymmetric QED with an FI-term, supports such strings~\DavisBS. In this theory they turn out to be BPS, with tension~$T_{\rm BPS} \sim \xi$.

Note that the real two-form~$F_{\mu\nu}$ in~$\chi_\alpha$ is proportional to the~$U(1)$ field strength in~$W_\alpha$. Since~$F_{\mu\nu}$ must be closed, we conclude that there are no magnetic charges in~$U(1)$ gauge theories with an FI-term.

\subsec{Chern-Simons Terms}

Consider a free~$U(1)$ gauge theory with a Chern-Simons term and an FI-term in three dimensions:
\eqn\CSlag{{\scr L} = -{1 \over 4 e^2} \int d^4 \theta \, \Sigma^2 + k \int d^4 \theta \, \Sigma V + \xi \int d^4 \theta \, V~.}
Here~$\Sigma = i \b D D V$ is the three-dimensional field strength; it is a real linear superfield. Using the equations of motion~$i \b D D \Sigma = 2 e^2 \xi + 4 e^2 k \Sigma~$, we find that the theory has an~$\CR$-multiplet
\eqn\CSrmult{\eqalign{& \CR_{\alpha\beta} = {1 \over 2 e^2} \left( D_\alpha \Sigma \, \b D_\beta \Sigma + D_\beta \Sigma \, \b D_\alpha \Sigma\right)~,\cr
& \chi_\alpha = i \b D_\alpha \CJ~, \qquad \CJ = -\xi \Sigma- {i \over 4 e^2} \b D D \left( \Sigma^2\right)~.}}
If~$\xi = 0$, we can perform an improvement transformation~\tdsmulit\ with~$U \sim {1 \over e^2} \Sigma^2$ to obtain an FZ-multiplet
\eqn\CSfzmult{\eqalign{& \CJ_{\alpha\beta} = {1 \over 2 e^2} \left( D_\alpha \Sigma \, \b D_\beta \Sigma + D_\beta \Sigma\, \b D_\alpha \Sigma\right) - {1 \over 16 e^2} \left( [ D_\alpha, \b D_\beta] + [ D_\beta, \b D_\alpha]\right) \left(\Sigma^2\right)~,\cr
& \CY_\alpha = D_\alpha X, \qquad X = {1 \over 16 e^2 } \b D^2 \left(\Sigma^2\right)~.}}
Note that the Chern-Simons level~$k$ does not appear explicitly.

\subsec{Real Mass Terms}

Three-dimensional~$\CN=2$ theories allow real mass terms. Each real mass parameter~$m$ is associated with a~$U(1)$ flavor symmetry. The flavor current is usually embedded in a real linear multiplet~$\CJ_m$, which contributes to the operator~$\chi_\alpha$ in the~$\CS$-multiplet,
\eqn\remchij{\chi_\alpha \sim i m \b D_\alpha \CJ_m~.}
Thus~$\chi_\alpha$ cannot be improved to zero in theories with real mass terms.

\subsec{Two-Dimensional $\CN=(0,2)$ K\"ahler Sigma Models}

Consider a two-dimensional~$\CN=(0,2)$ sigma model, whose target space is a K\"ahler manifold, such as~$\C\P^1$. The Lagrangian is
\eqn\hetsiglag{{\scr L} = {i\over 8} \int d\theta^+ d \thetabar^+ \, \d_i K \d_{--} \Phi^i + {\rm c.c.}~,}
where~$K$ is the K\"ahler potential and the~$\Phi^i$ are chiral,~$\b D_+ \Phi^i = 0$.  The classical theory is superconformal, and it admits an~$\CS$-multiplet~\hetsmult\ with~$\CS_{++} \sim g_{i \jbar} D_+ \Phi^i \b D_+ \b \Phi^{\jbar}$ and~$\CW_-=0$.  Quantum corrections lead to a breakdown of conformal invariance, and a non-zero~$\CW_-$ is generated at one-loop,
\eqn\oneloopw{\CW_- \sim R_{i \jbar} \, \d_{--} \Phi^i \b D_+ \bar\Phi^{\jbar}~.}
Here~$R_{i\jbar} = \d_i \d_{\jbar} \log \det g_{k \b l}$ is the Ricci tensor of the target space. This also shows that the~$R$-symmetry of the classical theory is anomalous. Note that we can write~$\CW_-=i\bar D_+ \CR_{--}$ with~$\CR_{--}\sim - i \partial_{i} \log\det g_{k\bar l} \, \partial_{--}\Phi^i$, which is not globally well-defined. Therefore, the~$R$-symmetry is not violated in perturbation theory, even though the theory does not admit a well-defined~$\CR$-multiplet. In particular, the constant~$C$ in~\hetsmult\ cannot be generated perturbatively.

Nonperturbatively, instantons activate the anomaly and explicitly break the~$R$-symmetry. To see this in more detail, let us consider the Euclidean two-point function $\langle S_{+--}(0) S_{+++} (z,\bar z)\rangle$,\foot{Here~$z, \b z$ are the Euclidean continuations of~$x^{--}, x^{++}$.} where~$S_{+--} \sim R_{i\bar j} \psi^i_+ \partial_{--}\b \phi^{\b j}$ is generated by the one-loop anomaly~\oneloopw\ and~$S_{+++} \sim g_{i \jbar} \psi_+^i \d_{++} \b \phi^{\jbar}$. Since this correlation function violates the~$R$-symmetry by two units, it vanishes in perturbation theory. However, instantons that violate the~$R$-symmetry by the same amount can lead to a nonzero answer.  For instance, this happens in the $\C\P^1$ model, where the (anti-) instanton of degree~$-1$ has two fermion zero modes, and thus gives rise to a contribution
\eqn\instcomp{\langle S_{+--}(0) S_{+++} (z,\bar z)\rangle_{\rm instanton} \sim {\Lambda^2 \over \b z}~.}
Here~$\Lambda$ is the strong coupling scale of the theory.  Upon integration,  the residue at~$\b z = 0$ gives rise to a contribution~$C\sim \Lambda^2$ in~\hetcurralg. We conclude that instantons generate the constant~$C$ in~\hetsmult. This was pointed out in~\WittenPX\ and explicitly verified in~\TanMI.

\subsec{A Quantum Mechanical Example}

An interesting class of examples in which the superfield~$\CY_\alpha$ in the~$\CS$-multiplet cannot be expressed in terms of a chiral superfield~$X$ consists of Wess-Zumino models whose superpotential~$W$ is not well-defined (see subsection~2.3). To briefly illustrate the interesting quantum effects that can arise in such models, we consider the~$\CN=2$ quantum mechanics of a real superfield~$\Phi$:
\eqn\qmlag{L = \int d \theta d \thetabar \left(\b D \Phi D\Phi + W(\Phi)\right)~.}
Here the superpotential~$W$ is real. Since we are interested in the case where~$W$ is not well-defined, we identify
\eqn\qmid{\Phi \sim \Phi + 2\pi~,}
and we choose
\eqn\qmw{W = f \Phi + \cos \Phi~,}
where~$f$ is a real constant. The classical vacua are determined by the equation
\eqn\clvac{\sin \Phi = f~.}
When~$|f| > 1$, there is no solution to~\clvac\ and SUSY is spontaneously broken at tree level. When~$0 < |f| < 1$, there are two classical supersymmetric vacua satisfying~\clvac.  In this case the system has two different instantons, which interpolate between these vacua -- one for each arc of the circle~\qmid.  These instantons mix the two vacuum states and lead to spontaneous SUSY-breaking.  Thus, the model~\qmlag\ spontaneously breaks SUSY for all non-zero values of~$f$. When~$f=0$, there are supersymmetric vacua at~$\Phi = 0, \pi$.  Now the two instantons are still present and each one mixes the two vacua, but their contributions exactly cancel and supersymmetry is unbroken.

Similar effects can arise in two-dimensional~$\CN=(2,2)$ theories when~$W$ is not well-defined. These models often admit BPS solitons that preserve some of the supercharges to all orders in perturbation theory. However, just as in the quantum mechanical example above, nonperturbative effects can break the remaining supersymmetries, so that the BPS property is not maintained in the full quantum theory~\refs{\HouRV,\BinosiWY}.

\newsec{Partial Supersymmetry Breaking and Space-Filling Branes}

The goal of this section is to clarify some issues about the phenomenon known as partial supersymmetry breaking, and to relate it to our previous discussion about supercurrent multiplets and brane currents.

\subsec{A Quantum Mechanical Example}

Following \HughesDN, we consider a quantum mechanical system with~$\CN=2$ supersymmetry
\eqn\qmssal{\eqalign{
&\{Q, \bar Q\}=2  H~,\cr
&\{Q, Q\}=2 Z~,\cr
&\{\bar Q, \bar Q\}=2 \bar Z~.}}
Here~$H$ is the Hamiltonian and the complex constant~$Z$ is a central charge. Note that the energy~$E$ satisfies the BPS bound~$E \geq |Z|$. Let us study the representations of the algebra~\qmssal\ as a function of~$Z$.

If $Z=0$, the algebra has two-dimensional representations with generic energy~$E>0$, and a one-dimensional representation with~$E=0$. The one-dimensional representation is supersymmetric; it is annihilated by both supercharges. If the Hilbert space includes a state in this representation, $\CN=2$ supersymmetry is unbroken. If there is no such state in the Hilbert space, supersymmetry is completely broken.

For~$Z \neq 0$, the situation is more interesting.  The representations with generic energy~$E>|Z|$ are two-dimensional, and they are similar to the two-dimensional representations of the~$Z=0$ algebra. In particular, both supercharges act non-trivially. There is also a one-dimensional representation with~$E=|Z|$, which saturates the BPS bound. It is annihilated by one linear combination of the supercharges, while the other linear combination acts as a constant. We say that such a state breaks~$\CN=2$ to~$\CN=1$. In other words, it partially breaks supersymmetry.

Virtually all models have a~$\Z_2$ symmetry, implemented by~$(-1)^F$, under which all fermions are odd.  Let us add this operator to the algebra~\qmssal.  Most of the representations discussed above easily accommodate this operator. The only exception is the one-dimensional representation with~$E=|Z|\not=0$, which must be extended to a two-dimensional representation.\foot{More generally, adding~$(-1)^F$ to the SUSY algebra doubles the size of a representation, whenever the number of supercharges that do not annihilate that representation is odd.}

There is a fundamental difference between the partial supersymmetry breaking that can happen when~$Z \neq 0$ and the spontaneous supersymmetry breaking that can happen when~$Z = 0$.

If~$Z=0$, the algebra~\qmssal\ admits supersymmetric representations. It is a dynamical question whether or not the Hilbert space of the system includes such supersymmetric states. Thus, whether or not supersymmetry is spontaneously broken is a property of the ground state. The high-energy behavior of the system is supersymmetric.

Turning on a non-zero~$Z$ does not spontaneously break~$\CN=2$ to $\CN=1$.  Instead, the original~$\CN=2$ supersymmetry algebra with~$Z=0$ is deformed.  This deformation of the algebra is a property of the high-energy theory rather than a property of the ground state.  The ground state is determined by the dynamics. If it saturates the BPS bound, $E=|Z|$, then~$\CN=1$ is preserved. If all states have~$E>|Z|$, supersymmetry is completely broken.

From this point of view, Witten's argument ruling out spontaneous partial supersymmetry breaking~\WittenNF\ is correct.  It applies to the algebra with~$Z=0$.  The observation of~\HughesDN\ is that the algebra can be deformed to admit states that partially break supersymmetry.

This quantum mechanical discussion also emphasizes the fact that partial supersymmetry breaking has nothing to do with infinite volume or with the non-existence of the supercharges. It is simply a consequence of deforming the supersymmetry algebra.

\subsec{Partial Supersymmetry Breaking and Space-Filling Branes}

In higher-dimensional systems, a central charge like~$Z$ in \qmssal\ is proportional to the volume of space.  For example, integrating the three-dimensional~$\CN=2$ current algebra~\tdsusyvar\ gives
\eqn\tdpsbalg{\{\b Q_\alpha, Q_\beta\} = -2 E A\gamma^0_{\alpha\beta} ~, \qquad \{Q_\alpha, Q_\beta\} =  {\b C A \over 4} \gamma^0_{\alpha\beta}~.}
Here~$E$ is the vacuum energy density, and~$A \rightarrow \infty$ is the spatial volume. This leads to the BPS bound~$E \geq |C| / 8$, so that the vacuum has positive energy whenever~$C \neq 0$. If the BPS bound is saturated, $E=|C|/8$, then the vacuum breaks half of the supercharges, preserving only~$\CN=1$ supersymmetry. If the BPS bound is not saturated, then SUSY is completely broken.

The same phenomenon occurs in two-dimensional~$\CN=(0,2)$ theories. Integrating the current algebra~\hetcurralg\ leads to
\eqn\twdpsbalg{\{ \b Q_\pm, Q_\pm\} = 2 E  L~, \qquad \{ Q_+, Q_+\} = -{\b C L \over 4}~,}
where~$L \rightarrow \infty$ is the spatial volume. Just as in three dimensions, we obtain a BPS bound~$E \geq |C|/8$. If this bound is saturated, then the vacuum preserves only one real supercharge and SUSY is partially broken from~$\CN=(0,2)$ to~$\CN=(0,1)$. Otherwise, supersymmetry is completely broken.

It should now be clear that partial supersymmetry breaking can be interpreted in terms of space-filling brane currents, which give rise to constants in the SUSY current algebra~\HughesDN. The deformation of the ordinary current algebra by these constants implies that some of the supersymmetries are always realized non-linearly.

\subsec{Examples}

There are copious known examples of theories with partial supersymmetry breaking.  Many of them arise as effective field theories on various BPS branes in field theory and string theory.

Perhaps the simplest examples occur in quantum mechanics.  The theory of D0-branes exhibits partial supersymmetry breaking.  This was used in the BFSS matrix model~\BanksVH\ and further explored in~\BanksNN.  Two-dimensional examples arise on the world sheet of strings.  The standard Green-Schwarz light-cone string exhibits~$\CN=16$ supersymmetry broken to~$\CN=8$.  Other examples in two dimensions were studied in~\refs{\HughesDN,\LosevGS}.

An interesting phenomenon arises in the two-dimensional~$\CN=(0,2)$ sigma model~\hetsiglag.  As we discussed in subsection~6.4, the constant~$C$ in~\twdpsbalg\ cannot be generated perturbatively.  Thus, the theory and its vacuum preserve~$\CN=(0,2)$ supersymmetry to all orders in perturbation theory.  Nonperturbatively, the constant~$C$ is generated by instantons, and the supersymmetry algebra is deformed~\refs{\WittenPX,\TanMI}. Therefore, the vacuum preserves at most one real supercharge.  As was pointed out in section~7.1, BPS vacua that preserve one real supercharge must come in pairs in order to represent~$(-1)^F$. Such pairs of BPS vacua do not constitute short representations, and consequently it is not easy to establish their existence.

Three-dimensional theories with partial supersymmetry breaking can be found on the world-volume of BPS domain walls embedded in four dimensions.  These theories admit a three-dimensional~$\CS$-multiplet~\tdsmult\ with~$C \sim z_{\rm BPS}$, which leads to partial SUSY-breaking~\AchucarroQB. In this class of models, the constants in the SUSY current algebra arise due to the presence of physical space-filling branes embedded in a higher-dimensional theory.

Another three-dimensional example is a variant of the two-dimensional model studied in~\LosevGS. It has a space-filling brane current at tree-level. We start with a Wess-Zumino model with a single chiral superfield~$\Phi$, canonical K\"ahler potential, and superpotential
\eqn\fdtdlag{W = \omega \log \Phi~.}
Here~$\omega$ is a complex constant. Note that this $W$ is not globally well-defined.  The model has a~$U(1)$ flavor symmetry under which~$\Phi$ has charge~$1$ and~$W$ is shifted by a constant,
\eqn\uonecharge{\eqalign{& \delta_{U(1)} \Phi = - i \Phi~,\cr
& \delta_{U(1)} W = -i\omega~.}}

The scalar potential leads to runaway behavior and the theory does not have a ground state. In order to avoid the runaway, we turn on a real mass~$m$ for the~$U(1)$ flavor symmetry. This stabilizes the runaway potential and it deforms the supercovariant derivatives by the action of the~$U(1)$ symmetry,
\eqn\newdops{\eqalign{& D_\alpha \rightarrow D_\alpha + m \thetabar_\alpha \delta_{U(1)}~,\cr
& \b D_\alpha \rightarrow \b D_\alpha + m \theta_\alpha \delta_{U(1)}~.}}
The chiral superfield~$\Phi$ still satisfies~$\b D_\alpha \Phi = 0$. Using the equations of motion~$\b D^2 \b \Phi = -{4 \omega \over \Phi}$, we can check that this model has an~$\CS$-multiplet~\tdsmult\ with
\eqn\LSmodeltd{\eqalign{& \CS_{\alpha\beta} = D_\alpha \Phi \b D_\beta \b \Phi + D_\beta \Phi \b D_\alpha \b \Phi~,\cr
& \chi_\alpha = - \half \b D^2 D_\alpha \left( \b \Phi\Phi\right) - 4 i m \b D_\alpha\left(\b\Phi \Phi\right)~,\cr
& \CY_\alpha = 4 \omega {D_\alpha \Phi \over \Phi}~,\cr
& C = 16 i m \omega~.}}
The vacuum saturates the BPS bound and supersymmetry is partially broken from~$\CN=2$ to~$\CN=1$.

The interpretation of partial supersymmetry breaking in terms of space-filling brane currents also applies to four-dimensional~$\CN=2$ theories. Examples of such theories are world-volume theories of BPS three-branes embedded in six dimensions~\HughesFA, and gauge theories with magnetic FI-terms~\refs{\AntoniadisVB, \FerraraXI}. At low energies, these models are described by four-dimensional Born-Infeld actions with~$\CN=1$ supersymmetry~\refs{\BaggerWP,\RocekHI}.

\newsec{Constraints on Renormalization Group Flow}

Consider a supersymmetric quantum field theory with a UV cutoff.  This theory must have a well-defined supercurrent multiplet.  In this section we discuss the behavior of this multiplet under renormalization group flow. This allows us to constrain the IR behavior of the theory.

All supercurrents furnish short representations of the supersymmetry algebra. (Equivalently, they satisfy certain constraints in superspace.)  As is typical in supersymmetric theories, short multiplets are protected: they must remain short under renormalization group flow. Therefore, the structure of the supercurrent multiplet is determined in the UV. This structure is then preserved at all energy scales along the renormalization group flow to the IR.

Before presenting specific applications of this reasoning, we would like to emphasize three important subtleties:
\smallskip
\item{1.)} In the extreme UV the theory is superconformal and it has a superconformal multiplet. As we start flowing toward the IR, the superconformal multiplet mixes with another multiplet and becomes larger -- it turns into one of the multiplets discussed above.  In this section, we would like to discuss the renormalization group flow starting at a high, but finite UV cutoff.
\medskip
\item{2.)} The opposite phenomenon happens in the extreme IR, where the theory is again superconformal and the multiplet becomes shorter.  This happens because some non-trivial operators flow to zero at the IR fixed point.  (If the low-energy theory is completely massive, the entire multiplet flows to zero in the extreme IR.) Therefore, our conclusions about the low-energy theory will be most interesting when we consider the theory at long, but finite distances.
\medskip
\item{3.)} The supercurrent multiplet must retain its form under renormalization group flow. In particular, constants that appear in the multiplet cannot change along the flow. This does not mean that these constants, or other operators in the multiplet, are not corrected in perturbation theory, or even nonperturbatively. However, these corrections are completely determined by the UV theory.

\bigskip

Consider a four-dimensional theory that admits an FZ-multiplet in the UV.  This FZ-multiplet must exist at all energy scales. Therefore, the theory cannot have strings carrying charge~$Z_\mu$. If the low-energy theory is a weakly coupled Wess-Zumino model, perhaps with some IR-free gauge fields, the existence of the FZ-multiplet in the IR implies that the target space of the Wess-Zumino model has an exact K\"ahler form (in particular, it cannot be compact), and that there is no FI-term for any~$U(1)$ gauge field~\refs{\KomargodskiPC,\KomargodskiRB}. This statement is nonperturbatively exact. It holds even if the topology of the target space or the emergence of~$U(1)$ gauge fields at low energies is the result of strong dynamics. (For earlier related results see~\WittenUP\ as referred to in~\AffleckVC, and~\refs{\ShifmanZI,\DineBK,\WeinbergUV}.) This reasoning can also be applied to constrain the dynamics of SUSY-breaking~\DumitrescuCA.

Likewise, a four-dimensional theory with a non-anomalous continuous~$R$-symmetry in the UV admits an~$\CR$-multiplet, and it must retain this multiplet at all energy scales. Consequently, a theory with a continuous~$R$-symmetry cannot support domain walls that carry charge~$Z_{\mu\nu}$. (Another application of tracking the~$\CR$-multiplet from the UV to the IR was recently found in~\AbelWV.) A theory that admits both an FZ-multiplet and an~$\CR$-multiplet supports neither strings nor domain walls with charges in the supersymmetry algebra.

Let us demonstrate this in specific examples. Pure~$SU(N_c)$ SUSY Yang-Mills theory admits an FZ-multiplet, but no~$\CR$-multiplet.\foot{The situation in this theory is similar to the discussion in subsection~6.4.  The superconformal invariance of the classical theory is broken by quantum corrections. At one-loop we find an FZ-multiplet with~$X \sim \Tr W^\alpha W_\alpha$, so that the~$R$-symmetry is anomalous. Even though the theory does not admit a well-defined~$\CR$-multiplet, the~$R$-symmetry is not violated in perturbation theory.  Nonperturbatively, instantons activate the anomaly and explicitly break the~$R$-symmetry.} It has~$N_c$ isolated vacua, and it supports domain walls carrying charge~$Z_{\mu\nu}$ that interpolate between these vacua~\DvaliXE. However, it does not support charged strings. On the other hand, SUSY QCD with~$N_f \geq N_c$ massless flavors has an FZ-multiplet and an~$\CR$-multiplet, and thus it supports neither strings carrying charge~$Z_\mu$ nor domain walls carrying charge~$Z_{\mu\nu}$.\foot{When~$N_f < N_c$, the theory does not have a stable vacuum and we do not discuss it~\AffleckMK.}  For~$N_c \leq N_f \leq {3 \over 2} N_c$, the IR theory is a weakly coupled Wess-Zumino model, in some cases with IR-free non-Abelian gauge fields~\refs{\SeibergBZ\SeibergPQ-\IntriligatorAU}. The target spaces of these Wess-Zumino models all have exact K\"ahler forms~\KomargodskiRB. This is particularly interesting in the case~$N_f = N_c$, when the topology of the IR target space is deformed~\SeibergBZ.

Just as in four dimensions, we can use supercurrents to constrain the IR behavior of supersymmetric field theories in two and three dimensions. In particular, we can establish whether a given theory admits branes, whose charges appear in the supersymmetry algebra. This is especially interesting for space-filling branes, which manifest themselves as constants in the various supercurrent multiplets. As such, they are not affected by renormalization group flow. If they are not present in the UV theory, they do not arise at low energies.

When comparing the UV and the IR theories, we must use the supercurrents of the full quantum theories. These may differ from the classical multiplets by perturbative or nonperturbative corrections. For example, we saw in subsection~6.4 that anomalies can modify the multiplet at one-loop. Likewise, the constant~$C$ in the two-dimensional~$\CS$-multiplet~\hetsmult\ can be be generated by instantons.  However, we emphasize again that this change in the value of~$C$ can be seen by performing an instanton computation in the UV theory.

One way to constrain the form of these quantum corrections is to follow~\SeibergVC\ and promote all coupling constants to background superfields.  For instance, we can introduce a coupling constant~$\tau$ in the sigma model~\hetsiglag\ by letting~$\d_i K \to \tau \d_i K$. We then promote~$\tau$ to a background superfield. It is clear from~\hetsmult\ that the constant~$C$ in the~$\CS$-multiplet can be modified by quantum corrections only if~$\tau$ is a chiral superfield, $\b D_+ \tau = 0$. This is the case for the~$\C\P^1$ model, since K\"ahler transformations in an instanton background force~$\tau$ to be chiral, and in this theory~$C$ is indeed generated by instantons~\refs{\WittenPX,\TanMI}. In sigma models whose target space has an exact K\"ahler form, $\tau$ can be promoted to an arbitrary complex superfield, and in this case~$C$ is not generated.\foot{$C$ violates the~$R$-symmetry by two units and therefore it can only be generated by instantons with two fermionic zero modes. Such instantons must be BPS, and they only exist in sigma models, whose K\"ahler form is not exact.  (See the related discussion around~\instcomp.)} (For a recent discussion of nonrenormalization theorems in two-dimensional~$\CN=(0,2)$ theories see~\CuiRZ.)

\vskip 1cm

\noindent {\bf Acknowledgments:}

We would like to thank T.~Banks, C.~Hull, D.~Jafferis, Z.~Komargodski, J.~Maldacena, M.~Rocek, M.~Shifman, W.~Siegel, A.~Vainshtein, and E.~Witten for
useful discussions. NS thanks the Simons Center for Geometry and Physics for its kind hospitality during the conclusion of this project. The work of TD was supported in part by a DOE Fellowship in High Energy Theory, NSF grant PHY-0756966, and a Centennial Fellowship from Princeton
University. The work of NS was supported in part by DOE grant
DE-FG02-90ER40542. Any opinions, findings, and
conclusions or recommendations expressed in this material are
those of the authors and do not necessarily reflect the views of
the National Science Foundation. {\it Note added in arXiv version 4:} We thank the authors of \Brunnerqyf\ for pointing out some typos corrected in this version of the present paper.

\appendix{A}{The Energy-Momentum Tensor}

In this appendix we review some facts about the energy-momentum tensor. Noether's theorem guarantees that any translation invariant local field theory possesses a real, conserved energy-momentum tensor~${\hat T}_{\mu\nu}$,
\eqn\conseq{\d^\nu {\hat T}_{\mu\nu} = 0~,}
which integrates to the total momentum
\eqn\mom{P_\mu = \int d^{D-1} x \;  {{\hat T}_\mu}^{\hskip4pt 0}~.}
The energy-momentum tensor is not unique. It can be modified by an improvement transformation,
\eqn\emimp{\hat T_{\mu\nu} \rightarrow \hat T_{\mu\nu} + \d^\rho B_{\mu\nu\rho}~,\qquad B_{\mu\nu\rho} = - B_{\mu\rho\nu}~.}
The improvement term~$\d^\rho B_{\mu\nu\rho}$ is automatically conserved and it does not contribute to the total momentum~\mom. For some choices of~$B_{\mu\nu\rho}$, the energy-momentum tensor~$\hat T_{\mu\nu}$ is not symmetric. (This is emphasized by the hat.) For instance, the canonical energy-momentum tensor in Lagrangian field theories is not symmetric, if the theory contains fields with non-zero spin.

Lorentz invariance guarantees that there is a choice for~$B_{\mu\nu\rho}$ that leads to a symmetric energy-momentum tensor~$T_{\mu\nu} = T_{\nu\mu}$. This is well-known for Lagrangian field theories~\BelinfantePU, but it holds more generally. Lorentz invariance implies the existence of a real conserved current~$j_{\mu\nu\rho}$,
\eqn\lorcurcons{\d^\rho j_{\mu\nu\rho} = 0~, \qquad j_{\mu\nu\rho} = - j_{\nu\mu\rho}~,}
which integrates to the Lorentz generators
\eqn\lgen{J_{\mu\nu} = \int d^{D-1}x \; {j_{\mu\nu}}^0~.}
The generators~$J_{\mu\nu}$ are time-independent and they satisfy~$i[P_\mu, J_{\nu\rho}] = \eta_{\mu\nu} P_\rho - \eta_{\mu\rho} P_\nu$, so that the current~$j_{\mu\nu\rho}$ must take the form
\eqn\lorcurr{j_{\mu\nu\rho} = x_\mu \hat T_{\nu\rho} - x_\nu \hat T_{\mu\rho} + s_{\mu\nu\rho}~, \qquad s_{\mu\nu\rho} = - s_{\nu\mu\rho}~.}
Here~$s_{\mu\nu\rho}$ is a local operator without explicit~$x$-dependence. We can obtain a symmetric energy-momentum tensor~$T_{\mu\nu}$ by performing an improvement transformation~\emimp\ with
\eqn\belinf{B_{\mu\nu\rho} = \half \left(s_{\nu\rho\mu} + s_{\nu\mu\rho} + s_{\mu\rho\nu}\right)~.}
In terms of~$T_{\mu\nu}$, the currents~\lorcurr\ can be written as~$j_{\mu\nu\rho} = x_\mu T_{\nu\rho} - x_\nu T_{\mu\rho}$, up to an overall improvement term.

The symmetric energy-momentum tensor~$T_{\mu\nu}$ is also not unique. It can be modified by further improvement transformations~\emimp, as long as~$B_{\mu\nu\rho}$ satisfies
\eqn\Bcons{\d^\rho B_{\mu\nu\rho} = \d^\rho B_{\nu\mu\rho}~,}
so that~$T_{\mu\nu}$ remains symmetric.\foot{Locally, we can express~$B_{\mu\nu\rho} = \d^\sigma Y_{\mu\sigma\nu\rho}$, where~$Y_{\mu\sigma\nu\rho}$ has the symmetries of the Riemann curvature tensor. (It is antisymmetric in each pair~$\mu\sigma$ and~$\nu\rho$, but symmetric under the exchange of these pairs). However, $Y_{\mu\sigma\nu\rho}$ may not be well-defined.} In general~$B_{\mu\nu\rho}$ has spin-1 and spin-2 components. If we restrict ourselves to the spin-1 component, we can write
\eqn\Bspone{B_{\mu\nu\rho}  = \eta_{\mu\rho} U_\nu - \eta_{\mu\nu} U_\rho~, \qquad \d_{[\mu} U_{\nu]} = 0~,}
so that the remaining allowed improvements for~$T_{\mu\nu}$ are given by
\eqn\Tspone{T_{\mu\nu} \rightarrow T_{\mu\nu} + \d_\mu U_\nu - \eta_{\mu\nu} \d^\rho U_\rho~.}
If there is a well-defined real scalar~$u$ such that~$U_\mu = \d_\mu u$, these improvements take the more familiar form
\eqn\Tsponii{T_{\mu\nu} \rightarrow T_{\mu\nu} + \left(\d_\mu \d_\nu - \eta_{\mu\nu} \d^2 \right) u~.}

\appendix{B}{Conventions in Two and Three Dimensions}

In this appendix we summarize our conventions for spinors and supersymmetry in two and three dimensions. Whenever possible, we use the dimensionally reduced conventions of Wess and Bagger~\WessCP. In any number~$D$ of dimensions, we take the Minkowski metric to be~$\eta_{\mu\nu} = - + \cdots +$, where the Lorentz indices~$\mu, \nu$ run from~$0$ to~$D-1$. We normalize the totally antisymmetric Levi-Civita symbol as~$\ep_{0 1 \cdots (D-1)} = -1$.

\subsec{Conventions in Three Dimensions}

In~$D=3$, the Lorentz group is~$SL(2, \R)$ and the fundamental representation is a real two-component spinor~$\psi_\alpha = \b \psi_\alpha \; \left(\alpha = 1,2\right)$. There are only undotted indices and as in~$D=4$, they are raised and lowered by acting from the left with the antisymmetric symbols~$\ep_{\alpha\beta}$ and~$\ep^{\alpha\beta}$,
\eqn\raislower{\psi^\alpha = \ep^{\alpha\beta} \psi_\beta~, \qquad \psi_\alpha = \ep_{\alpha\beta} \psi^\beta~.}
There is now only one way to suppress contracted spinor indices,
\eqn\contr{\psi\chi = \psi^\alpha \chi_\alpha~,}
and this leads to some unfamiliar signs, which are absent in~$D=4$. For instance, under Hermitian conjugation we have
\eqn\hc{\b {\left(\psi \chi\right)} = - \b \chi \b \psi~.}
We work in a basis in which the three-dimensional gamma matrices are given by\foot{These gamma-matrices are obtained by reducing~$\sigma^\mu_{\alpha\betadot}$ along the four-dimensional~2-direction.}
\eqn\gammamat{\gamma^\mu_{\alpha\beta} = \left(- \1, \sigma^1, \sigma^3\right)~.}
Here~$\1$ is the~$2 \times 2$ unit matrix, and~$\sigma^1, \sigma^3$ are Pauli matrices. The gamma matrices~\gammamat\ are real, and they satisfy the following identities:
\eqn\gamaid{\eqalign{& \gamma^\mu_{\alpha\beta} = \gamma^\mu_{\beta\alpha}~, \cr
& {\left(\gamma^\mu\right)_\alpha}^\beta {\left(\gamma^\nu\right)_\beta}^\lambda = \eta^{\mu\nu} {\delta_\alpha}^\lambda + \ep^{\mu\nu\rho} {\left(\gamma_\rho\right)_\alpha}^\lambda~,\cr
& \left(\gamma^\mu\right)_{\alpha\beta} \left(\gamma_\mu\right)_{\lambda\kappa} = \ep_{\alpha\lambda} \ep_{\kappa \beta} + \ep_{\alpha\kappa} \ep_{\lambda\beta}~.}}
We can use these to switch between vectors and symmetric bi-spinors,
\eqn\tdvbsp{\ell_{\alpha\beta} =  - 2 \gamma^\mu_{\alpha\beta} \ell_\mu~, \qquad \ell_\mu = {1 \over 4} \gamma^{\alpha\beta}_\mu \ell_{\alpha\beta}~,\qquad  \ell_{\alpha\beta} = \ell_{\beta\alpha}~.}

The conventional~$\CN=2$ supersymmetry algebra in~$D=3$ takes the form
\eqn\tdsalg{\eqalign{& \{Q_\alpha, \b Q_\beta\} = 2 \gamma^\mu_{\alpha\beta} P_\mu + 2 i \ep_{\alpha\beta}Z~,\cr
& \{Q_\alpha, Q_\beta\} = 0~.}}
The real scalar~$Z$ is a central charge. (As in four dimensions, we can extend~\tdsalg\ by adding additional brane charges~\FerraraTX.) This algebra admits a~$U(1)_R$ automorphism under which~$Q_\alpha$ has charge~$-1$,
\eqn\tdrch{[R, Q_\alpha] = - Q_\alpha~.}
If the central charge~$Z$ in~\tdsalg\ vanishes, then $\CN=2$ superspace in~$D=3$ is the naive dimensional reduction of~$\CN=1$ superspace in~$D=4$.  The supercharges~$Q_\alpha$ are represented on superfields~$S(x, \theta, \thetabar)$ by differential operators~$\CQ_\alpha$,
\eqn\chdiffiop{[\xi^\alpha Q_\alpha - \b \xi^\alpha \b Q_\alpha, S] = i \left(\xi^\alpha \CQ_\alpha - \b \xi^\alpha \b \CQ_\alpha\right) S~,}
with
\eqn\tdqdiffop{\eqalign{& \CQ_\alpha = {\d \over \d \theta^\alpha} + i \left(\gamma^\mu \thetabar\right)_\alpha \d_\mu~,\cr
& \b \CQ_\alpha = - { \d \over \d \thetabar^\alpha} - i \left(\gamma^\mu\theta\right)_\alpha \d_\mu~.}}
The corresponding supercovariant derivatives are given by
\eqn\tdcovder{\eqalign{& D_\alpha = {\d \over \d \theta^\alpha} - i \left(\gamma^\mu \thetabar\right)_\alpha \d_\mu~,\cr
& \b D_\alpha = - { \d \over \d \thetabar^\alpha} + i \left(\gamma^\mu\theta\right)_\alpha \d_\mu~.}}
They satisfy the identities
\eqn\Did{\eqalign{& \{ D_\alpha, \b D_\beta\} = i \d_{\alpha\beta}~, \cr
&  D^\alpha \b D_\alpha = \b D^\alpha D_\alpha~,\cr
& \{D_\alpha, D_\beta\} = 0~.}}
These formulas can be used to derive other useful identities, such as~$\b D_\alpha \b D^\beta D_\beta = - \half \b D^\beta \b D_\beta D_\alpha$. To write supersymmetric actions, we also need the superspace integrals
\eqn\tdssint{\int d^2 \theta \, \theta^2 = 1~, \qquad  \int d^2 \thetabar \, \thetabar^2 = -1~,\qquad \int d^4 \theta \, \theta^2 \b \theta^2 = - 1~.}

\subsec{Conventions in Two Dimensions}

In~$D=2$, the irreducible representations of the Lorentz group are real and one-dimensional. There are two inequivalent real spinors $\psi_\pm$. They can be obtained by reducing from~$D=3$ and identifying\foot{In our conventions, this corresponds to reducing along the three-dimensional~1-direction.}
\eqn\spinred{\psi_{\alpha = 1} \rightarrow \psi_-~, \qquad \psi_{\alpha = 2} \rightarrow \psi_+~.}
As in~\raislower, we raise and lower indices according to
\eqn\raislowertwd{\psi^+ = - \psi_-~, \qquad \psi^- = \psi_+~.}
We will only use spinor indices~$\pm$, so that every vector~$\ell_\mu$ is written as a bi-spinor
\eqn\tdvbsp{\ell_{\pm\pm} =  \ell^{\mp\mp} = 2 \left(\ell_0 \pm \ell_1\right)~.}
This leads to some unfamiliar numerical factors. For instance,
\eqn\ellsq{\ell^2 = - {1 \over 8} \left(\ell^{++} \ell_{++} + \ell^{--}\ell_{--}\right) = - {1 \over 4} \ell_{++} \ell_{--}~.}

The conventional~$\CN=(2,2)$ supersymmetry algebra in~$D=2$ takes the form
\eqn\twodsalg{\eqalign{& \{Q_\pm, \b Q_\pm\} = - P_{\pm\pm}~,\cr
& \{Q_+, Q_-\} = Z~,\cr
& \{ Q_+, \b Q_-\} = \t Z~.}}
The complex scalars~$Z$ and~$\t Z$ are central charges. This algebra admits a continuous~$U(1)_{R_V}\times U(1)_{R_A} $ automorphism
\eqn\rvect{\eqalign{
&[R_V, Q_\pm] = - Q_\pm~, \qquad [R_V, Z] = -2 Z~, \cr
&[R_A, Q_\pm] = \mp Q_\pm~,\qquad [R_A, \t Z] = -2 \t Z~,}}
as well as a~$\Z_2$ mirror automorphism
\eqn\mirroraut{Q_- \leftrightarrow \b Q_-~, \qquad Z \leftrightarrow \t Z~, \qquad R_V \leftrightarrow R_A~.}
If the central charges in~\twodsalg\ vanish, then~$\CN=(2,2)$ superspace in~$D=2$ is the naive dimensional reduction of~$\CN=2$ superspace in~$D=3$. The supercharges~$Q_\pm$ are represented on superfields~$S(x, \theta, \thetabar)$ by differential operators~$\CQ_\pm$,
\eqn\chdiffiop{[\xi^+ Q_+ + \xi^- Q_- - \b \xi^+ \b Q_+ - \b \xi^- \b Q_- , S] = i \left(\xi^+ \CQ_+ + \xi^- \CQ_- - \b \xi^+ \b \CQ_+ - \b \xi^- \b \CQ_-\right) S~,}
with
\eqn\towdqdiffop{\eqalign{& \CQ_\pm = {\d \over \d \theta^\pm} + {i \over 2} \thetabar^\pm \d_{\pm\pm}~,\cr
& \b \CQ_\pm = - { \d \over \d \thetabar^\pm} - {i\over 2}\theta^\pm \d_{\pm\pm}~.}}
The corresponding supercovariant derivatives are given by
\eqn\tdcovder{\eqalign{& D_\pm = {\d \over \d \theta^\pm} -{i \over 2} \thetabar^\pm \d_{\pm\pm}~,\cr
& \b D_\pm = - { \d \over \d \thetabar^\pm} + {i\over 2}\theta^\pm \d_{\pm\pm}~.}}
They satisfy the identities
\eqn\Did{\eqalign{& \{ D_\pm, \b D_\pm\} = i \d_{\pm\pm}~, \cr
& D_\pm^2 = \b D_\pm^2 = \{D_\pm, D_\mp\} = 0~.}}

The~$\CN=(0,2)$ subalgebra of~\twodsalg\ takes the from
\eqn\hetsalg{\eqalign{& \{ Q_+, \b Q_+\} = - P_{++}~,\cr
& Q_+^2 = 0~.}}
It admits a~$U(1)_R$ automorphism under which~$Q_+$ has charge~$-1$. To obtain~$\CN = (0,2)$ superspace, we simply set~$\theta^- = 0$ in~$\CN = (2,2)$ superspace.

\appendix{C}{The~$\CS$-Multiplet in Two-Dimensional~$\CN=(2,2)$ Theories}

The~$\CS$-multiplet in two-dimensional theories with~$\CN=(2,2)$ supersymmetry consists of two real superfields~$\CS_{\pm\pm}$, which satisfy the constraints
\eqn\twdsmult{\b D_\pm \CS_{\mp\mp} = \pm\left(\chi_\mp + \CY_\mp\right)~,}
where
\eqn\twdchicons{\eqalign{& \b D_\pm \chi_\pm = 0~,\cr
& \b D_\pm \chi_\mp = \pm C^{(\pm)}~,\cr
& D_+ \chi_- - \b D_- \b \chi_+ = k~,}}
and
\eqn\twdycons{\eqalign{& D_\pm \CY_\pm = 0~, \cr
& \b D_\pm \CY_\mp = \mp C^{(\pm)}~,\cr
& D_+ \CY_- + D_- \CY_+ = k'~.}}
Here~$k, k'$ and~$C^{(\pm)}$ are real and complex constants respectively.

It is straightforward to solve the constraints~\twdsmult\ in components:
\eqn\twdsmultcomp{\eqalign{\eqalign{\CS_{\pm\pm}  = ~& j_{\pm\pm} - i \theta^\pm S_{\pm\pm\pm} - i \theta^\mp \left( S_{\mp\pm\pm} \mp 2 \sqrt 2 i \b \psi_\pm\right) - i \thetabar^\pm \b S_{\pm\pm\pm} \cr
& - i \thetabar^\mp \left(\b S_{\mp\pm\pm} \pm 2 \sqrt2 i \psi_\pm\right) - \theta^\pm \thetabar^\pm T_{\pm\pm\pm\pm} + \theta^\mp \thetabar^\mp \left(A \mp { k + k' \over 2}\right) \cr
& + i \theta^+ \theta^- \b Y_{\pm\pm} + i \thetabar^+ \thetabar^- Y_{\pm\pm} \pm i \theta^+ \thetabar^- \b G_{\pm\pm} \mp i \theta^- \thetabar^+ G_{\pm\pm} \cr
& \mp \half \theta^+ \theta^- \thetabar^\pm \d_{\pm\pm} S_{\mp\pm\pm} \mp \half \theta^+ \theta^- \thetabar^\mp \d_{\pm\pm}\left(S_{\pm\mp\mp} \pm 2 \sqrt 2 i \b \psi_\mp\right)\cr
& \mp \half \thetabar^+ \thetabar^- \theta^\pm \d_{\pm\pm} \b S_{\mp\pm\pm} \mp \half \thetabar^+ \thetabar^-\theta^\mp \d_{\pm\pm} \left( \b S_{\pm\mp\mp} \mp 2 \sqrt 2  i \psi_\mp\right)\cr
& + {1 \over 4} \theta^+ \theta^- \thetabar^+ \thetabar^- \d_{\pm\pm}^2 j_{\mp\mp}~.}}}
The chiral superfields~$\chi_\pm$ are given by
\eqn\twdchicomp{\eqalign{&\chi_+ = - i \lambda_+(y) - i \theta^+ \b G_{++}(y) +  \theta^- \left(E(y) + {k \over 2}\right) + \thetabar^- C^{(-)}  + \theta^+ \theta^- \d_{++} \b \lambda_-(y)~,\cr
& \chi_- = - i \lambda_-(y) - \theta^+ \left(\b E(y) - {k \over 2}\right) + i \theta^- G_{--}(y) - \thetabar^+ C^{(+)}- \theta^+ \theta^- \d_{--} \b \lambda_+(y)~,\cr
& \lambda_\pm = \pm\b S_{\mp\pm\pm} + \sqrt 2 i \psi_\pm~,\cr
&E = \half \left( T_{++--} - A \right)+ {i \over 4} \left( \d_{++} j_{--} - \d_{--} j_{++}\right)~,\cr
& \d_{++} G_{--} - \d_{--} G_{++} = 0~,\cr
&y^{\pm\pm} = x^{\pm\pm} + 4 i  \theta^\pm\thetabar^\pm~,}}
and the twisted (anti-) chiral superfields~$\CY_\pm$ are given by
\eqn\twdycomp{\eqalign{&\CY_+ = \sqrt 2 \psi_+ (\b {\t y}) +  \theta^- \left( F(\b {\t y}) + {k' \over 2}\right) - i \thetabar^+ Y_{++}(\b {\t y}) - \thetabar^- C^{(-)}+ \sqrt2 i \theta^- \b \theta^+ \d_{++} \psi_-(\b {\t y})~,\cr
& \CY_- = \sqrt 2 \psi_-(\t y) -  \theta^+ \left( F(\t y) - {k' \over 2}\right) + \thetabar^+ C^{(+)}- i \thetabar^- Y_{--}(\t y) + \sqrt 2 i \theta^+ \thetabar^- \d_{--} \psi_+ (\t y)~,\cr
& F = - \half \left( T_{++--} + A\right) - {i \over 4} \left(\d_{++} j_{--} + \d_{--} j_{++}\right)~,\cr
& \d_{++} Y_{--} - \d_{--} Y_{++} = 0~,\cr
&\t y^{\pm\pm} = x^{\pm\pm} \pm 4 i \theta^\pm \thetabar^\pm~.}}
The supersymmetry current is conserved, and the energy-momentum tensor is real, conserved, and symmetric. The~$\CS$-multiplet contains~$8+8$ independent real operators and the constants~$k, k', C^{(\pm)}$.

The mirror automorphism~\mirroraut\ acts on superspace by exchanging~$\theta^- \leftrightarrow - \thetabar^-$ and $D_- \leftrightarrow \b D_-$. The constraints~\twdsmult, \twdchicons, and~\twdycons\ are invariant, if we accompany this action on superspace by
\eqn\smultmirror{\eqalign{& \CS_{\pm\pm} \leftrightarrow \pm \CS_{\pm\pm}~, \qquad \chi_+ \leftrightarrow \b \CY_+~, \qquad \chi_- \leftrightarrow - \CY_-~, \cr
& k \leftrightarrow -k'~, \qquad C^{(+)} \leftrightarrow C^{(+)}~, \qquad C^{(-)} \leftrightarrow \b C^{(-)}~.}}
This implies that~$Q_- \leftrightarrow \b Q_-$,~$Z \leftrightarrow \t Z$, and~$R_V \leftrightarrow R_A$.

The~$\CS$-multiplet of~$\CN = (2,2)$ decomposes into multiplets of the~$\CN=(0,2)$ subalgebra. This decomposition includes the~$\CS$-multiplet of $\CN=(0,2)$, which is given by
\eqn\hetred{\eqalign{& \CS_{++}\big|_{\theta^- = 0}~,\cr
& \CW_- = i \left(\chi_- - \CY_-\right)\big|_{\theta^- = 0}~,\cr
& \CT_{----} = \half [\b D_-, D_-] S_{--} \big|_{\theta^-=0}~.}}
These superfields satisfy the constraints~\hetsmult\ with~$C=2 i C^{(+)}$.  After the~$\CN=(0,2)$ projection, the constants~$k,k'$ can be eliminated by a shift of~$T_{++--}$, which amounts to an unobservable shift in the total energy.

The~$\CS$-multiplet~\twdsmult\ can be modified by an improvement transformation
\eqn\twdsmulit{\eqalign{& \CS_{\pm\pm} \rightarrow \CS_{\pm\pm} + [D_\pm, \b D_\pm] U ~,\cr
& \chi_\pm \rightarrow \chi_\pm - \b D_+ \b D_- D_\pm U~,\cr
& \CY_\pm \rightarrow \CY_\pm - D_\pm \b D_+ \b D_- U~.}}
Here~$U$ is a real superfield, which is well-defined up to shifts by a real constant.

In some cases, the~$\CS$-multiplet can be improved to a smaller supercurrent:
\smallskip
\item{1.)} If~$k = C^{(\pm)} = 0$ and there is a well-defined real~$U$ such that~$\chi_\pm = \b D_+ \b D_- D_\pm U$, then~$\chi_\pm$ can be improved to zero and we obtain an FZ-multiplet
    \eqn\twdfz{\eqalign{&\b D_\pm \CJ_{\mp\mp} = \pm\CY_\mp~,\cr
    & D_\pm \CY_\pm = 0~, \qquad \b D_\pm \CY_\mp = 0~,\cr
    & D_+ \CY_- + D_- \CY_+ = k'~.}}
    This multiplet contains~$4+4$ independent real operators and the real constant~$k'$. It follows from~\twdfz\ that
    \eqn\racons{\d_{++} \CJ_{--} - \d_{--} \CJ_{++} = 0~,}
    so that the bottom component of the FZ-multiplet gives rise to a conserved~$R_A$-current with~$R_A = -{1 \over 4} \int dx \left(j_{++} - j_{--}\right)$.
\medskip
\item{2.)} If~$k' = C^{(\pm)} = 0$ and there is a well-defined real~$U$ such that~$\CY_\pm = D_\pm \b D_+ \b D_- U$, then~$\CY_\pm$ can be improved to zero and we obtain an~$\CR$-multiplet
    \eqn\twdrmult{\eqalign{& \b D_\pm \CR_{\mp\mp} = \pm\chi_\mp~,\cr
    & \b D_\pm \chi_+ = 0~, \qquad \b D_\pm \chi_- = 0~,\cr
    & D_+ \chi_- - \b D_- \b \chi_+ = k~.}}
    Like the FZ-multiplet, it contains~$4+4$ real operators, as well as the real constant~$k$. It follows from~\twdrmult\ that
    \eqn\rvcons{\d_{++} \CR_{--} + \d_{--} \CR_{++} = 0~,}
    so that the bottom component of the~$\CR$-multiplet is a conserved~$R_V$-current with $R_V = -{1 \over 4} \int dx \left(j_{++} + j_{--}\right)$.  Note that the mirror automorphism~\mirroraut\ exchanges the~$\CR$-multiplet and the FZ-multiplet.
\medskip
\item{3.)} If~$k = k' = C^{(\pm)} = 0$ and we can set both~$\chi_\pm$ and~$\CY_\pm$ to zero by a single improvement transformation, then the theory is superconformal and the supercurrent satisfies
    \eqn\twdscft{\b D_\pm \CJ_{++} = 0~, \qquad \b D_\pm \CJ_{--} = 0~.}
\bigskip

The current algebra that follows from the~$\CS$-multiplet takes the form
\eqn\twdsusyvar{\eqalign{& \{ \b Q_\pm, S_{\pm\pm\pm}\} = - T_{\pm\pm\pm\pm} - { i \over 2} \d_{\pm\pm} j_{\pm\pm}~,\cr
& \{\b Q_\pm, S_{\pm\mp\mp} \} = - T_{++--} \pm \half\left(k - k'\right) + { i \over 2} \d_{\mp\mp} j_{\pm\pm}~,\cr
& \{Q_\pm, S_{\pm\mp\mp}\} = 2 \b C^{(\pm)}~,\cr
& \{Q_+, S_{-\pm\pm}\} = \mp i \b Y_{\pm\pm}~,\cr
& \{\b Q_+, S_{-\pm\pm}\} = \mp i G_{\pm\pm}~.}}
This allows us to identify the conserved brane currents. The zero-brane currents~$\mp i \b Y_{\pm\pm}$ and~$\pm i \b G_{\pm\pm}$ give rise to well-defined central charges~$Z$ and~$\t Z$. These currents are exchanged by the mirror automorphism~\mirroraut.  The constants~$C^{(\pm)}$ and~$k - k'$ are interpreted as space-filling~brane currents, which can lead to partial SUSY-breaking.

\appendix{D}{Additional Supercurrent Multiplets?}

In this appendix we consider certain multiplets that are more general than the $\CS$-multiplet.  We show that they are acceptable supercurrents only if they differ from the $\CS$-multiplet by an improvement transformation.

One such multiplet was proposed in~\refs{\MagroAJ,\KuzenkoAM}; see also~\refs{\ZhengXX,\KuzenkoNI}. It is a real superfield~$\CK_\mu$ that satisfies the constraints
\eqn\ttmult{\eqalign{& \b D^\alphadot \CK_{\alpha\alphadot} = \chi_\alpha + i \chi'_\alpha + \CY_\alpha~,\cr
& \b D_\alphadot \chi_\alpha = 0~,\qquad D^\alpha \chi_\alpha = \b D_\alphadot \b \chi^\alphadot~,\cr
& \b D_\alphadot \chi'_\alpha = 0~,\qquad D^\alpha \chi'_\alpha = \b D_\alphadot \b \chi'^\alphadot~,\cr
& D_\alpha\CY_\beta + D_\beta \CY_\alpha = 0~, \qquad \b D^2 \CY_\alpha = 0~.}}
If~$\chi'_\alpha = 0$, we recover the~$\CS$-multiplet~\fdsmult.

It is straightforward to solve the constraints~\ttmult\ in components. In particular, we find that
\eqn\ttmltcomp{\CK_\mu\big|_{\theta \sigma^\nu \thetabar} =  2 \hat T_{\nu\mu} - \eta_{\mu\nu} A - {1 \over 8} \ep_{\nu\mu\rho\sigma} \left( F^{\rho\sigma} + 4 \d^\rho j^\sigma\right) - \half F'_{\nu\mu}~.}
The operators~$A, F_{\mu\nu}, j_\mu$ are familiar from the~$\CS$-multiplet, while the real closed two-form~$F'_{\mu\nu}$ comes from the superfield~$\chi'_\alpha$. The energy-momentum tensor~$\hat T_{\mu\nu}$ is real and conserved,
\eqn\conseq{\d^\nu \hat T_{\mu\nu} = 0~,}
but it is~{\it not} symmetric,
\eqn\emas{\hat T_{\mu\nu} - \hat T_{\nu\mu} = {1 \over 4} F'_{\mu\nu}~.}
However, Lorentz invariance guarantees that~$\hat T_{\mu\nu}$ can be improved to a symmetric energy-momentum tensor~$T_{\mu\nu}$ (see appendix~A).

The allowed improvements of~\ttmult\ take the form
\eqn\ttimp{\eqalign{& \CK_{\alpha\alphadot} \rightarrow \CK_{\alpha\alphadot} + D_\alpha \b \Sigma_\alphadot - \b D_\alphadot \Sigma_\alpha~,\cr
& \chi_\alpha \rightarrow \chi_\alpha + {3 \over 4} \left( \b D^2 \Sigma_\alpha - 2 \b D_\alphadot D_\alpha \b \Sigma^\alphadot - D_\alpha \b D_\alphadot \b \Sigma^\alphadot\right)~,\cr
& \chi'_\alpha \rightarrow \chi'_\alpha - {i \over 4} \left( \b D^2 \Sigma_\alpha + 2 \b D_\alphadot D_\alpha \b \Sigma^\alphadot + D_\alpha \b D_\alphadot \b \Sigma^\alphadot\right)~,\cr
& \CY_\alpha \rightarrow \CY_\alpha + \half D_\alpha \b D_\alphadot \b \Sigma^\alphadot~,}}
where~$\Sigma_\alpha$ satisfies the constraint
\eqn\omegacons{D_\alpha \Sigma_\beta + D_\beta \Sigma_\alpha = 0~.}
The transformation~\ttimp\ shifts the energy-momentum tensor by an improvement term of the form~\symimp,
\eqn\thatbelin{\hat T_{\nu\mu} \rightarrow \hat T_{\nu\mu} + \d_\nu U_\mu - \eta_{\mu\nu} \d^\rho U_\rho~, \qquad \d_{[\mu} U_{\nu]} = 0~.}
If this improvement makes~$\hat T_{\mu\nu}$ symmetric, then~\emas\ shows that it also sets the two-form~$F'_{\mu\nu}$ to zero, and hence the entire superfield~$\chi'_\alpha$ vanishes. Thus, the multiplet~\ttmult\ is an acceptable supercurrent only if it is decomposable and can be improved to an~$\CS$-multiplet.

As an example, we consider a supercurrent that arises in conjunction with non-minimal supergravity theories~\refs{\GatesYC,\GatesNR}. In our conventions it takes the form
\eqn\GGRSmulta{\eqalign{
    &\b D^\alphadot \CG_{\alpha\alphadot} =  i \chi'_\alpha + D_\alpha X ~,    \cr
    &\chi'_\alpha = -{i\over 6}\left(\b D^2 \lambda_\alpha + 2 \b D_\alphadot D_\alpha \b \lambda^\alphadot +  D_\alpha \b D_\alphadot \b \lambda^\alphadot \right)~, \cr
    &X=  {1 \over 3( 3n + 1)} \b D_\alphadot \b \lambda^\alphadot~,}}
where
\eqn\lambdacon{D_\alpha \lambda_\beta + D_\beta \lambda_\alpha=0~,}
and~$n$ is a complex parameter. We immediately see that it is possible to set~$\chi'_\alpha$ to zero by an improvement transformation~\ttimp\ with~$\Sigma_\alpha = -{2\over 3} \lambda_\alpha$. This gives rise to an~$\CS$-multiplet
\eqn\ggrsmulaimp{ \CS_{\alpha\alphadot} = \CG_{\alpha\alphadot} -{2\over 3} \left( D_\alpha \bar \lambda_\alphadot - \bar D_\alphadot \lambda_\alpha\right)~,}
with
\eqn\newG{\eqalign{&\chi_\alpha = -{1\over 2}\left(\b D^2 \lambda_\alpha - 2 \b D_\alphadot D_\alpha \b \lambda^\alphadot - D_\alpha \b D_\alphadot \b \lambda^\alphadot \right)~, \cr
& X=  -{n \over  3n + 1} \b D_\alphadot \b \lambda^\alphadot~.}}
This form of the multiplet makes manifest two special cases:~if~$n=0$, we obtain an~$\CR$-multiplet, and if~$n \rightarrow -{1\over 3}$, we obtain an FZ-multiplet. (These values of~$n$ correspond to the new-minimal and the old-minimal limits of non-minimal supergravity.) The~$\CS$-multiplet~\ggrsmulaimp\ is decomposable when~$\lambda_\alpha = D_\alpha U$, where the real superfield~$U$ is well-defined up to shifts by a real constant.  Then, it can be improved to either an FZ-multiplet or an~$\CR$-multiplet. In particular, in this case the theory has a continuous~$R$-symmetry.

\listrefs

\end